\newcommand{\onepic}[2]{\includegraphics[scale=#1]{#2}}
\documentclass[10pt]{article}

\usepackage{amsmath}
\usepackage{amssymb}

\usepackage{graphicx}

\usepackage{cite}

\usepackage{color}

\usepackage[labelfont=bf,labelsep=period,justification=raggedright]{caption}

\makeatletter
\renewcommand{\@biblabel}[1]{\quad#1.}
\makeatother

\date{}

\begin{document}

\begin{flushleft}
{\Large
\textbf{Scaling in the global spreading patterns of pandemic Influenza A (H1N1) and the role of control: empirical statistics and modeling}
}
\\
Xiao-Pu Han $^{1,2}$,
Bing-Hong Wang $^{1,3}$,
Changsong Zhou $^{2,4,\ast}$,
Tao Zhou  $^{1,5}$,
Jun-Fang Zhu  $^{1}$
\\
\bf{1} Department of Modern Physics, University of Science and Technology of China, Hefei 230026 China
\\
\bf{2} Department of Physics, Hong Kong Baptist University, Hong Kong
\\
\bf{3} The Research Center for Complex System Science, University of Shanghai for Science and Technology and
Shanghai Academy of System Science, Shanghai, 200093 China
\\
\bf{4} Centre for Nonlinear Studies, and The Beijing-Hong Kong-Singapore Joint Centre for Nonlinear and Complex Systems (Hong Kong),  Hong Kong
\\
\bf{5} Web Sciences Center, University of Electronic Science and Technology of China, Chengdu 610054, China

$\ast$ E-mail: Corresponding author:  cszhou@hkbu.edu.hk
\end{flushleft}

\section*{Abstract}
{\it \bf  Background}: The pandemic of influenza A (H1N1) is a serious on-going global
public crisis. Understanding its spreading dynamics is of
fundamental importance for both public health and scientific
researches.  Recent studies have focused mainly on evaluation and prediction of on-going spreading,
which strongly depends on detailed information about the structure of social contacts, human traveling patterns and biological activity of virus, etc.

\noindent
{\it \bf Methodology/Principal Findings}:
In this work we analyzed the distributions of
confirmed cases of influenza A (H1N1) in different levels and find the Zipf's law and Heaps' law.
Similar scaling properties were also observed for severe acute respiratory syndrome (SARS) and bird
cases of H5N1. We also found  a hierarchical spreading pattern from countries with larger population and
GDP to countries with smaller ones. We proposed a  model that considers  generic control effects on both the local
growth and transregional transmission,  without the need of the above mentioned detailed information.
We studied in detail the impact of control effects  and heterogeneity on the spreading dynamics in the model and showed 
that they  are responsible for the scaling and hierarchical spreading properties observed in empirical data.

\noindent
{\it \bf  Conclusions/Significance}:
Our analysis and modeling showed that although strict control measures for interregional
travelers could delay the outbreak in the regions without
local cases, the focus should be turned to local prevention after the outbreak of local cases.
Target control on a few regions with the  largest number of active interregional travelers  can efficiently prevent the spreading.
This work provided not only a deeper understanding of the generic mechanisms underlying the spread of infectious diseases,
but also some practical guidelines for
 decision makers to adopt suitable control strategies.

\section*{Introduction}

A new global influenza pandemic has broken out. In the
first three months, the epidemic spreaded to over 130 countries, and
more than $10^5$ people were infected by the novel virus
influenza A (H1N1). H1N1 represents a
very serious threat due to cross-species transmissibility and the
risk of mutation to new virus with increased transmissibility.
Several early studies paid attention to this public issue from
different perspectives \cite{Neumann,Smith,Rohani,Fraser,Coburn,Balcan2009,YangY,Wallinga}, and
made known important information such as the biological activity of
H1N1 virus and the patterns of early spreading. While
every effort was taken to develop antiviral and vaccination
drugs, efficient reduction of the spreading could have
already been achieved by interventions of population contact. However,
such interventions, like strict physical checking at the borders and
enforced quarantine, are costly and highly controversial. It is
therefore difficult to decide the control strategies: when should
the schools be suspended and whether the border control should be
reinforced or given up?


The detailed mechanism of transmission can differ significantly for
different virus, the spreading patterns, however, may display common
regularities due to generic contacting processes and control
schemes. Many health organizations have collected large amount of
information about the spreading of H1N1. In-depth analysis of these
data, together with what we have known for SARS \cite{Huf,Masu}, avian influenza (H5N1)
\cite{Small,Colizza2007}, foot-and-mouth epidemic \cite{Ferg1,Ferg2}
and some other pandemic influenza \cite{Ferg0,Longi}, may lead us to
a more comprehensive understanding of the common spreading patterns that do
not rely on the detailed biological features of virus. In this
paper, we studied the spreading patterns of influenza pandemic by both
empirical analysis and modeling. Our main contributions were
threefold: (i) The Zipf's law of the distribution of confirmed cases
in different regions were observed in the spreading of H1N1, SARS and
H5N1; (ii) A simple model was proposed, which does not rely on the
biological details but can reproduce the observed scaling
properties; (iii) The significant effects of control strategies were
highlighted: the strong control for interregional travel is
responsible for the Zipf's law and can sharply delay the
outbreak in the regions without local cases, while the focus should
be turned to local prevention after the outbreak of local cases. Our
analysis provided a deeper understanding of the relationship between
control and spreading, which is very meaningful for decision makers.

\section*{Results}

\subsection*{Empirical Results}

We first analyzed the {\it cumulative} number $n_i$ of laboratory confirmed
cases of H1N1 of each country $i$ to a given date (see the data description in
{\it Materials and Methods}). Because $n_i$ is growing, the
distributions for different dates are normalized by the global total
cases $N_T=\sum n_i$ to  the corresponding dates for comparison.
What we used in our analysis is the Zipf's plots \cite{Zipf}, which was obtained by sorting  each $n_i>0$ in a descending order,
from rank 1 to the largest value $M$ and plotting $n_i$ with respect to  the rank $r_i$.
We considered the normalized Zipf's plot where  each  $n_i$ was replaced by its corresponding proportion $P_i = n_i/N_T$.  More discussion of  the Zipf's plot can be found in {\it Materials and Methods}.
Table 1 shows the ranking of the  top 20 countries and their total of confirmed cases in  five typical dates.
Fig.1(a) and 1(b) report the Zipf's plots for the distributions of normalized $n_i$ in different
dates. The maximal rank  $M$ corresponds to the number of regions with
confirmed cases, which grows during the spreading.
The normalized distributions $P$ surprisingly
display scaling properties. Before the middle of  May, $P$ shows clearly a
power-law type $P\sim r^{-\alpha}$  with an exponent $\alpha$ changing around
$3.0$ (except the first data point, see Fig. 1(a)). Although the total
cases $N_T$ grows rapidly in this early stage, $P$ for different
dates seems to follow the same line in the log-log plot. After the
middle of May, the middle part of the distribution grows more
quickly, and meanwhile the virus spreads quickly to many more
countries. 
In this stage the exponent $\alpha$ of the left part of $P$ steadily reduces from higher than $3.0$ to $1.7$ 
(see Fig. S3(b) in {\it Supporting Information}), and an exponential tail emerges (Fig. 1(b)). After June 10, $P$ can be well fitted by a power-law
function with an exponential tail, for example, $P \sim r^{-1.70}e^{-0.013r}$ for the data of July 6 (solid line, Fig. 1 (b)).
The scaling properties are not special for H1N1, but quite common in various diseases, such as SARS in
2003  (Fig. 1(c), $\alpha \approx 2.7$) and the bird cases of H5N1 in 2008 (Fig. 1(d), $\alpha \approx 2.0$),
although  the spreading range is much more limited (to only about 30 countries).

There could be variations or errors in the real-world surveillance of H1N1, which  may affect  the ranking of the  countries. To examine the robustness of our analysis
against such variations, we considered several types of  possible variation: (A)  the variation is correlated with the reported total of cases in a country; (B) the variation is correlated with the population of the country;  and
(C)  the variation is correlated with both the reported total of cases and the population of the country.
We found that  the form of power-law-like distribution in Zipf's plot is robust under these different types of variations (Fig. S1).This analysis shows that our finding of  the power-law-like form  is still believable
in the presence of  variations or errors in surveillance. The detailed discussion can be found in the \emph{Supporting Information}.

Another scaling property which is often accompanied by Zipf's law is the {Heaps' law} \cite{Heaps,Zhang2008,Cattuto2009}.
Heaps' law describes a sublinear growth of the number of distinct sets  $M (t) $ as the increasing
 of the total number of elements $N_T(t)$ belonging to those sets, with the power -law form $M \sim N_T^{\lambda}$.
 A detailed introduction can be found in {\it Materials and Methods}. In the pandemic of H1N1,
 the number of infected countries $M$ and the global total confirmed cases $N_T$ obeys the Heaps' law with the Heaps' exponent $\lambda\approx 0.35$ before
 May 18, 2009 and $\lambda \approx 0.53$ after May 18 (Fig. 2). The exponents of the Zipf's law and the Heaps law
 satisfy  $\alpha \lambda \approx 1 $, which is consistent with the theoretical analysis \cite{Lu2009} that if an evolving system has a stable
Zipf's exponent, its growth must obey the Heaps' law with exponent $\lambda\approx 1/\alpha$ (an approximate
estimation when $\alpha>1$: the larger the $\alpha$, the more accurate the estimation).After May 18, the pronounced exponential tail in the distribution $P$ (Fig. 1(b)) leads to a deviation from strict Zipf's law,
and the two exponents no longer satisfy the relationship $\lambda\approx 1/\alpha$.

We also found that broad distribution of  $n_i$ is related to
heterogeneity in different countries.  Figs. 3(a)
and 3(b) report the dependence between the number of confirmed cases $n_i$
and the population and gross domestic product (GDP). A clearly
hierarchical spreading pattern, similar to what were predicted by
some theoretical complex network models \cite{Bar,Yang}, can be observed: the big
and rich countries were infected first, and then the disease spread  out to the global world. This can be understood that
bigger and richer countries usually have more active population in
international travel, and thus are of higher risk to be new spreading
origins in the early stage of epidemic.
The evolution of correlations between the confirmed cases $n_i$ and population and GDP was
reported in Fig. 3(c) by the \emph{Kendall's Tau} $\tau_K$.
Kendall's Tau measures the correlations between two datesets which are strongly heterogeneous in magnitude.
The method to calculate $\tau_K$ was introduced in {\it Materials and Methods}.
To test whether $\tau_K$ is significant, we compared the value from original datasets to those from
surrogate data by shuffling the order of the population of GDP (see {\it Supporting Information}).
Significantly positive correlations with a tendency of increase with time can be observed (Fig. 3(c) and Fig. S2).
These results show that the interconnectivity among world regions and human mobility are important
factors that accelerate the spread of diseases globally.
The global total confirmed cases $N_T$ displays two
phases of growth (Fig. 3(d)): in the early stage $N_T$ increases with a high rate and then turns into a stable
exponential growth with a much smaller rate, with the transition occurring around the
middle of May. Such a transition  may reflect the changes in the
contacting rate among people due to imposed or self-adaptive control.

It is interesting to study whether and how the exponent $\alpha$ in the distribution $P$ is related to the well-known reporductive
number $R$ in mathematical epidemic theory.
Employing the method in Ref. \cite{Li} and the estimate of the mean serial interval of 3.2 days in Ref. \cite{Cowl0,Cowl1},
we estimated, using the growth rate in the stable growth period after the middle of May,  that $R$ is between 1.09 and 1.22
for the serial interval in the range [1.9, 4.5] days.
This range is slightly smaller than the results in several other estimations based on the early spreading \cite{Fraser,Ni},
but generally in agreement with other studies based on the spreading after the early outbreak \cite{Cowl1}.
As seen in Fig. S3(a) in {\it Supporting Information}, there is a rapid decrease of $R$ in the period before the middle of May.
As we will show later in the model, the decrease of $R$ could be attributed to the control effects.
Interestingly,
the power-law exponent $\alpha$ of the distribution $P$  has a similar trend of evolution (Fig. S3(b)).
An  positive correlations can be found in the plot $R$ vs. $\alpha$ for the early stage when both of them are relative large.
This relationship could be explained as follows. In the early stage of spreading, $R$ is effectively higher when surveillance and control schemes
for H1N1 were not very effective. On the other hand,  only  a small portion of the population was infected. In this stage, local growth was quicker than transmission between countries,
so that the reported number decreases quickly with rank,
  corresponding to a larger $\alpha$. 
  However, a simple relationship between $R$ and $\alpha$ is not expected because $\alpha$ from the normalized
  distribution of a given date  is related to the accumulated effects of $R$ before this date, especially in the later stage of spreading. 


We also investigated the statistical regularities within a country.
We compared the normalized distribution of confirmed cases
in different states of USA and in different provinces of China (Fig. 4):
$P$ of USA shows a much more homogeneous form with a large deviation
from strict power-law distribution while $P$ of China is close to a
power-law with exponent $\alpha=1.79 \pm 0.04$ (the Zipf's distribution of
SARS cases of different provinces of China is also a power-law type
with exponent $\alpha \approx 3$ \cite{Wu}).


We have investigated the growth of the number of confirmed cases $n_i$  for all the 13 countries with $n_i>10^3$ until July 6,
and found that the patterns are quite diverse.
As shown in Fig. 5, some have a clear transition in the middle of May from a rapid breakout to a stably exponential
growth (e.g. U.S.A. and Mexico), which is similar to the global growth patterns;
some have much later initial infections (e.g. Australia); some exhibit a stably exponential growth without
a pronounced crossover (e.g. China); and some show irregular growth curves (e.g. Japan and New Zealand).
The spreading of H1N1 was impacted by many factors, such as control measures, traffic systems, school terms, and so on,
which could lead to such diverse growth patterns under a stable global growth.

To summarize, the empirical results show that the scaling properties in epidemic spreading process may widely exist at different regional levels and crossing various infectious diseases. 
In the following we tried to obtain some insight into the generic mechanisms underlying these common properties.

\subsection*{Modeling and Simulation Results}

The empirical results provoke some outstanding questions: how to
understand the scaling properties in region distributions, which
factors lead to the different spreading patterns for different
regions, and what are the effects of control measures on the
regional level spreading? We believed that the scaling properties
have the origin at the generic contact process underlying the
transmission of diseases, and the variation could result from the
heterogeneity of the contact process in different diseases and
regions. One most important heterogeneity may be the control
strength. To build a generic model incorporating the effects of
control, let us consider the actions taken by people when facing a
serious epidemic spreading. In general, individual people try to
take many approaches to reduce the probability of infection, such as
using respirator, reducing the face-to-face social interactions, and
disinfecting frequently. Meanwhile, many organizations usually take
measures to prevent the spreading of epidemic, such as physical
examinations in public transportation and schools,
isolation for highly risky groups, and so on. If epidemic breaks out in
a country, other countries may reinforce the health examinations at
the borders for the travelers from that country. For example, in China,
measurement of body temperature was used in many airports and border crossings, and
the identified infected persons and their close contacts were strictly isolated in the early stage of H1N1 spreading.
In Hong Kong, students had to measure body temperature and were not allowed to go to school
when the temperature was higher than a threshold.
These actions of individuals and social
organizations can effectively change the structure of social
contacts, reduce infection probability and affect the spreading
patterns of epidemic \cite{Gross,Han}. Such effect of imposed or self-adaptive controlling
actions was the starting point of our model.


Different from many individual-based models, our model is in the
regional level, so the detailed social contact structure
\cite{Moore,Newm,Pastor0,Moreno,Moreno1} as well as the control methods and
strategies in individual level \cite{Pastor,Huerta,Cohen} are not
considered directly.
Our scheme was based on the metapopulation framework. In this framework,
the global community is divided into a set of regions, each having its own spreading dynamics, but also interacting with each other.
This framework has been widely used in modeling epidemic spreading in the last decade \cite{Gren,Keel,Fulf,Watt,Verg}.
In our model, a region (such as a country) is denoted by a node in a network with $K$ nodes in total.
Different from previous work considering details of transportation \cite{Huf, net1} or mobility \cite{net2} networks,
the network is supposed to be fully-connected since in general there
are direct contacts between almost all countries in the world.
However, the strength of connections between countries could be
different due to the heterogeneity in various factors, such as population and economics.
As will be shown later, while such heterogeneity has some
impact on the epidemic spreading, the most important ingredients are
the strengths of control within and between regions. Therefore,
instead of employing the detailed information of real traffics, we
generically denoted the international traffic of a node as its
strength $s_i$, and the weight of link between two node $i$ and $j$
is assumed to be symmetric and proportional to the products of the
strengths $s_i$ and $s_j$:
\begin{equation}
q_{ij} = s_{i}s_{j}/{\sum}_{k=1}^{K}s_{k}.
\end{equation}

The spreading at time $t$ from node $j$ to $i$ is proportional to
the number of infected cases $n_j$ of node $j$, together with a
time-varying effective weight $w_{ij}(t)$ of the link, namely
$w_{ij}(t)n_j(t)$. Here $w_{ij}(t)$ is related not only to the link
strength $q_{ij}$, but also to the control strategy. Control measures
are in general reinforced on the travelers from countries with large
number of infected cases, and thus in our model the link weight is
\begin{equation}
w_{ij}(t) = q_{ij}n_{j}(t)^{-\beta _{1}},
\end{equation}
where $\beta_1$ is a free parameter. Effectively, we can take
$w_{ij}(t)=0$ if $n_j(t)=0$. Note that while $q_{ij}$ is symmetric,
$w_{ij}$ is in general asymmetric. This expression describes
generically the effects of various control measures at the borders,
without relying on the details at the individual level.


In this model, the update of the number of cases $n_{i}$ of an
arbitrary node $i$ consists of two parts: a local infection growth
and the global traveling infections:
\begin{equation}
\Delta n_i = \rho \left[a_{i}(t) n_{i}(t) + \frac{b}{\langle s
\rangle}{\sum}_{j = 1, j\neq i}^{K}w_{ij}(t)n_j(t) \right],
\end{equation}
where $\rho$ is a positive constant related to the basic
transmissibility of the diseases, $\langle s \rangle$, the average
value of $s_{i}$, is introduced for normalization, and the
coefficient $b$ denotes the relative contribution due to the
transmission from other regions. Note that $\Delta n$ is generally a
real number while the real-world increment of infected cases must be
integral. Therefore, we round $\Delta n$ to the neighboring integer,
namely to set $n_i(t+1)=n_i(t)+ [\Delta n]+1$ with probability $p$
and $n_i(t+1)=n_i(t)+[\Delta n]$ with probability $1 - p$, where $p
= \Delta n - [\Delta n]$ ($[x]$ denotes the largest integer no
larger than $x$).

The relative contribution by local infections, $a_{i}(t)$, is not
constant, but reflects the strength of control within a region. In
the same vein as the border control in Eq. 2, we described the
generic effects of local control by decaying $a_i(t)$ as a function
of $n_i$ with a free parameter $\beta_2$, namely
\begin{equation}
a_{i}(t) =
\left\{
    \begin{array}{cc}
n_{i}(t)^{-\beta _{2}}, &$if $ n_{i}(t)^{-\beta _{2}} > g \\
g, &$if $ n_{i}(t)^{-\beta _{2}} {\leq} g.
    \end{array}
    \right.
\end{equation}
Effectively, $a_i(t)=0$ if $n_i=0$. Here the decaying of $a_i$ is
limited by a constant $g$ ($0 < g < 1$), which accounts for
the necessary social contacts in the daily life even under the outbreak of the epidemic.
In reality, $g$ is also related to the transmissibility and death rate of the disease.

In our model, $n_i$ is the total infected cases of a node. In reality, the reported and confirmed cases
are most likely a small part of the total cases. If we assume that the ratio of reported  cases is similar
for different countries and roughly constant in time, the model can be used to describe the
distribution of confirmed cases without changing our conclusions in the following.

To focus on the effects of the control parameters $\beta_1$ and $\beta_2$, we first considered  the simplest case in which $s_{i}$ is uniform.
In this case, $q_{ij}=1/K $ and Eq. (3) is reduced to the minimal model
\begin{equation}
\Delta n_i = \rho \left[ a_{i}(t) n_{i}(t) + \frac{b}{K}{\sum}_{j = 1, j\neq i}^{K}n_j(t)^{1-\beta_1} \right ].
\end{equation}
The impact of the heterogeneity in $s_i$ will be discussed later.

We would like to emphasize that the effect of control considered in our model does not refer in particular to any of the specific control measures.
Eqs. (2) and (4) are supposed to describe generically the integrated effects
of various intervention schemes, either imposed by govermental policies or self-organized by individuals.
 For example, the border control parameter $\beta_1$ describes the integrated effect of all the measures impacting on the spreading across different countries, and
$\beta_2$ not only includes the impact of some official control measures, but also the impact of some adaptive individual actions,
such as reducing of social contact, and wearing gauze mask.  All of our discussions of "control" are based on this extended meaning.

In the following, to represent a worldwide network of countries,
the total number of nodes in our model is $K=220$.
The model can also be used to represent the spreading within a county when regarding a node as a region within a country and
ignoring the transmission from other countries.
From Eq. (3), the parameter $\rho$ does not affect the pattern of the normalized distribution $P$. $\rho$
is thus fixed at $0.2$ in all our simulations, which is close to the fast growing rate of the influenza A in the early stage of outbreak (see Fig. 3(d)).
We quantified the epidemic spreading initiated randomly at one node by  the spreading range $M$ (the number of nodes with $n_i>0$) and the total cases $N_T$ 
and investigate how they depend on the control parameters $\beta_1$ and $\beta_2$ (Fig. 6).
It is seen that
both large $\beta_1$ and large $\beta_2$ can reduce the range of spreading $M$, but the control on the interregional borders by $\beta_1$ is more effective
than $\beta_2$ (Fig. 6(a)). On the contrary, large $\beta_2$ is much more effective than $\beta_1$ to reduce the total number of cases  $N_T$ (Fig. 6(b)).
The patterns in Fig. 6 are generic in the model for different parameters $\rho$, $b$ and $g$ and for
different time during the spreading. These results imply that once a country has local epidemic outbreaks, its growth will be mainly driven by the local spreading but not the input of foreign cases.

The parameter space of $\beta_1$ and $\beta_2$ can be divided into four regimes, corresponding to the combinations of weak or strong
and local or interregional controls, as indicated  in Fig. 6. Typical normalized  distributions $P$ obtained in the four regimes are compared in Fig. 7.
When $\beta_1$ is small (regimes (I) and (II)), the  epidemics  can spread to almost all nodes in short time, and $P$ is rather homogeneous.
When $\beta_1$ is large (regimes (III) and (IV)), the spreading across different region is suppressed, and $P$ is rather inhomogeneous,
manifested as a power-law-like form. Keeping $\beta_2 $ fixed,  the exponent $\alpha$ clearly increases with $\beta_1$  and a larger $\beta_2$
 can slightly increase $\alpha$ further  (inset, Fig. 7(a)). We have included a detailed discussion of the time evolution of the distributions $P$
 and their association to the Heaps' law in Fig. S5  of  the \emph{Supporting Information}.

While $\beta_1$ has a sensitive impact on the interregional spreading and controls the heterogeneity of the distribution $P$, $\beta_2$ mainly affects the growth of total cases $N_T$, especially in the early stage (Fig. 7(b)).  
With stronger control at larger $\beta_2$, the fast growth of $N_T$ in the early stage will be effectively suppressed and transformed to a slow exponential growth within shorter time. As seen in  Eq. (4), $\beta_2$ only affects the growth of the epidemic in the very early stage after it appears in a region. The significant effect of $\beta_2$ on the growth of total cases $N_T$ emphasizes the importance of early epidemic control, in agreement with the conclusion of previous studies on other diseases \cite{Ferg1, Ferg2}.

Comparing  the results from the four regimes, we can see that the spreading pattern in regime III (large $\beta_1$ and small $\beta_2$)  is closer to the empirical observations of influenza A. In this regime,  the range of $\alpha$ covers most of empirical results. For example, with  $\beta_1 = 0.8$ and $\beta_2 = 0.2$, the distribution $P$ can be well fitted by a power-law function with exponent $1.67$ (Fig. 7(a)), and this value is close to the empirical exponent of influenza A  on July 6 (Fig. 1(b)). Large $\beta_1$ and small $\beta_2$ is consistent with the real-world situation. While more efficient to implement control measures on the borders, e.g.,  to identify infected and suspected candidates and their close contacts for quarantine, it is much more difficult to get the same efficiency for the same control schemes  in  local communities. The relative lower death rate of influenza A is also likely to weaken the self-adaptive control and voluntary isolation of the individuals, leading to insignificant change of the contact patterns  (e.g., much weaker than SARS). All these will render a lower efficiency in the local control, corresponding to a small $\beta_2$ and a larger $g$.


Besides the two parameters $\beta_1$ and $\beta_2$ for the border and local control, the other two parameters $b$ and $g$ related to interregional and local contact rates can also significantly affect the spreading processes. The parameter $b$ in our model denotes the relative strength of interregional transmission. Large flow of interregional travels can also make the epidemic spread to most of the regions rapidly. As a result, the distribution $P$ becomes more homogeneous with decreasing $\alpha$ when $b$ is larger; however, $b$ has only a slight  impact  on the growth pattern of $N_T$ (Fig.  8(a)).

The parameter $g$ expresses the background local growth speed which cannot be further reduced due to unavoidable social contacts even under the effect control measures.  Under strong border control (large $\beta_1$), the number of  infected cases  $n_i$ is mainly determined by $g$, growing exponentially with the rate $\rho g$ after an initial transient period, thus $g$ has a very sensitive impact on the growth of the total number $N_T$ (Fig. 8(b)). If $g$ is large, earlier infected regions  will have much more infected cases compared to later infected regions, leading to an inhomogeneous distribution $P$. At smaller $g$, the earlier and later infected regions do not differ very much in the number of infected cases, corresponding to more homogeneous distribution $P$ with decreasing $\alpha$ (Fig. 8(b)). Different from the case of weak border control (small $\beta_1$), homogeneous $P$ here dose not mean the rapid spreading; on the contrary, it denotes the situation that the infection in each country is in a low level.


All the above discussions are based on the minimal model where the diversity of the nodes and the edges is ignored by assuming a uniform $s_i$. Now we study the impact of heterogeneous $s_i$ and the effect of target control on the spreading of disease.  While  previous investigations have focused overwhelmingly on the impact of  heterogeneity in the degree of complex networks \cite{Bar,Pastor0,Moreno, Moreno1}, here we study  the effects of the heterogeneity in the intensity of  nodes and links in globally coupled networks. We first took the real population of different countries as $s_i$ in our model and investigate how does the initiation of the disease in countries with different ranks of populations influence the global spreading. When the disease starts in a country with a large $s_i$ (Fig. 9(a), population rank $R_{ini} = 11$ as Mexico), the disease spreads out quickly and
the  spreading process displays a clear tendency from the node with large $s_i$ to those  with small $s_i$ as seen by the evolution of the scatter plot of $n_i(t)$ vs. $s_i$ and the Kendall's tau (Fig. 9(c)), which reproduces the main features in the empirical data in Fig. 3. On the contrary, when the initiation happens in a country with small population  (Fig. 9(b), population rank $R_{ini} = 100$ as Libya), the disease is contained in the country where it is initiated for a period of time, and then
the countries with the largest populations get infected soon and become new centers of spreading. $\tau_K$ is around zero in the very beginning when the diseases is contained and becomes negative when spreading to a few nodes with the largest $s_i$  and
quickly shift to positive values when the new centers take the leading role in the spreading Fig. 9(c)). The total cases $N_T$ grows much faster in the first case (Fig. 9(d)). We  applied target control in our model (see {\it Material and Methods}), and we found that strong control just on one or two nodes with the largest $s_i$ can sharply reduce the spreading by several orders of magnitude (Fig. 9(e)).  This effect is similar to target immunization of the hubs in degree heterogeneous complex networks \cite{Pastor}. Here the results are shown for one realization of the simulation. The statistics over many
realizations displayed in Fig. S6 in the {\it Supporting Information} can evidently demonstrate the spreading from the nodes with large $s_i$ to those with small $s_i$.
A more systematic analysis of the effects of node heterogeneity and target control by considering a power-law distribution of $s_i$ is included in
the {\it Supporting Information}. We find that even though the heterogeneity can accelerate the spreading (Fig. S7),
the strength of control plays the leading role to determine the patterns of spreading (Fig. S8).
The spreading can be sharply decelerated by reducing both the total cases $N_T$ and range $M$, when only a few nodes with the largest $s_i$ are in
strong border control  (Fig. S9).

Another extension of our model considers the diverse effects of control in different country to qualitatively explain the different growth patterns for different country shown in Fig. 5.
We assume that the parameters $\beta_1$ and $\beta_2$  are nonidentical and are randomly chosen between $0$ and  $1$ for different nodes, while we fix the other parameters. In reality, all the important parameters $\rho$, $\beta_1$, $\beta_2$, $g$, and $b$  can be different due to variation of contact structures (population, hygiene condition, culture, etc.) from country to country. Here we did not intend to fit the model precisely to the real data, but rather to demonstrate the concept and to prove the principle.

The results were summarized in Fig. 10 for two groups of nodes with early and late initial infections. In each group, we consider four combinations of the parameters $\beta_1$ and $\beta_2$. For the nodes where the disease is initiated and got infected in the very beginning (Fig. 10 (a)), the other nodes are not infected and there is no significant input, thus  the growth patterns are dominantly determined by $\beta_2$.  When $\beta_2$  is close to 1, the local growth rate shifts quickly to $\rho g$, corresponding to an exponential function without a clear transition. A pronounced transition happens when $\beta_2$ is close to zero and it takes a period time for the local growth rate to settle down to $\rho g$. The growth patterns of the late infected nodes  depend on both $\beta_1$ and $\beta_2$ (Fig. 10(b)): the transition to stably exponential growth is still determined by $\beta_2$, while larger  $\beta_1$ prevents the input from other nodes and makes $n_i$ smaller. In all the cases, the stable growth rates are close to $\rho g$ (the slope $\approx \rho g \log_{10} e$),  therefore additional variation of $\rho g$ can account for diverse exponential growth rates in the data. We can see that the basic growth patterns in empirical data, i.e., with and without a pronounced transition, can be reproduced by different control parameters in the model. The model, however, does not include strong non-stationary ingredients that could lead to sudden increase of $n_i$ observed in a few countries in Fig. 5.

Fig. 10 also shows the corresponding $N_T$ in this model of diverse control parameters, which also reproduce the feature of $N_T$ in the data.
The behavior is similar if we further include the diversity in the parameter $g$.  We would like to point out that the growth patterns of $n_i$
in the individual nodes with different parameters are similar to various growth patterns of the global total $N_T$ in the model
without diversity in parameters.  This provides justification that we can apply our model to the global level where each node represents a country, or to the level within a country where each node denotes a state/province. In the later case, $N_T$ of the model represents the growth of the total cases of a country and is consistent with the growth of $n_i$ in the former case when similar interregional parameters $\beta_2$ and $g$ are considered in both level, namely the model at different level will provide consistent conclusion about the epidemic spreading.

To summarize, power-law distribution of $P$ with large exponent $\alpha$ appears in situations with large $\beta_1$, small $b$ and large $g$. This regime corresponds to the real situations that the epidemic control for the travelers is strong, the interregional contact is much weaker compared to that in local communities, and the change of local social contacts by the disease  is not very significant. The epidemic control for the interregional travelers (large $\beta_1$) is the most important condition for the emergence of the power-law type of $P$, since the power-law distribution cannot be generated when $\beta_1$ is close to zero no matter what other parameters are.

\section*{Discussion}

The statistics of region distributions of several pandemic diseases, including H1N1,
SARS and bird cases of H5N1 display obvious scaling properties in the spreading process at different levels.
We studied the origin of such scaling properties
with a model of epidemic spreading at the regional level that incorporates the generic effects of
intervention and control measures without the need of the structure details of social contacts and the particularity of the transmission of the diseases.
Such a model is then able to capture the general principles underlying epidemic spreading and to reveal the generic impact of  control measures.
We elucidated that strict epidemic control on interregional travellers plays an important role in the emergence of the scaling properties.

The results of the model can cover the empirical statistics of  H1N1 on both the region distribution
and the growth of total cases, and  are also consistent with the region distributions of SARS and H5N1.
 In particular,  the exponent  $\alpha$ of the empirical  distribution $P$
of H1N1 is about $3.0$ in the early stage and changes to  $1.7$ on July 6, 2009, and $\alpha$ is about
$2.7$ for SARS and $2.0$ for H5N1. In the stable spreading period, the
$\alpha$ of  H1N1 is smaller than SARS and H5N1. According to the understanding from our model, larger
$\alpha$  indicates that  the control measures  are more strict and effective. This  is in agreement
with the situation in SARS and H5N1 spreading. Because of
high death rate and strong infection capability, SARS gave  rise to strong social panic and attracted
attentions from citizens to governments in the countries with outbreaks, such as China, and
strict  control measures  were enforced in each public transportation systems and in daily life of
people. As for H5N1, many efficient control measures were also taken to  prevent the spreading, such as immunity for poultry and
culling of livestock, etc.
Large $\alpha$ in the early stage of the spreading of H1N1 could be related to stronger
control effect due to overrating of  the mortality of H1N1. Empirical results also showed that
the distribution $P$ of H1N1 in USA is more homogeneous than in  China. While there are probably several factors contributing to
 this difference, but the most obvious difference is in the control measures.  China took strict control policies,
 such as entry screening at airports and border crossings, and enhanced surveillance of outpatients and inpatients with influenza-like illness, enforced quarantine and isolation for identified infected persons and the close contacts, which are not so strict compared to those during the SARS spreading, but are stronger than USA.


Our main findings, i.e., interregional control  mainly affects  the spreading range and the form of the region distributions while local control sensitively impacts  the growth of total cases,  provide us a  picture of epidemic control.  For regions that have no or only a few local infected persons, strict control measures for interregional  travellers can delay the local outbreaks significantly, but if there are large number of local cases, these control methods for travellers are not so important.  Instead, control methods and treatment for local communities will be much more  helpful. After the Summer of 2009,  the focal point of the control policies for H1N1 of many countries turned to the treatment for infected persons. According to the conclusions of the present model, this strategy shift  is reasonable. This model also indicates that  the diversity of different regions will accelerate the spreading. Efficient prevention of  the spreading could be achieved by enhanced control measures, especially for the giant regions. Further work will be focusing on the impact of target  \cite{Wallinga} or voluntary vaccination \cite{Haifeng}.

In summary, a simple physical model basing on the abstraction of the generic contact processing and the effects of control
can provide meaningful understanding of the scaling properties commonly observed in various pandemic diseases.
 It deepens our understanding of the relationship between the strength of control and the spreading process, and provides a
 meaningful guidance  for the decision maker to adopt suitable control strategies.

\section*{Materials and Methods}
\subsection{Data Description}
The  cumulative number of laboratory confirmed
cases of H1N1 of each country is available from the website of
Epidemic and Pandemic Alert of World Health Organization (WHO)
(http://www.who.int/), which started
from April 26 to July 6, and updated each one or two days. Each update is in a new webpage, for example, the data in May 21 is shown in the webpape (http://www.who.int/csr/don/2009\_05\_21/en/ index.html). After
July 6, WHO stopped the update for each country since the global
pandemic has broken out.

Table 1 lists the countries with  the rank of confirmed cases up to 20 in several typical dates.
The corresponding total of confirmed cases is shown after the country name in the table.
A complete list of the data in these typical dates can be found in \emph{Supporting  Information}. 

The data for SARS and H5N1 are respectively available from the websites of WHO (http://www. who.int/csr/sars/country/en/index.html) and the World Organization for Animal
Health (OIE)(http: //www.oie.int/wahis/public.php?page=disease) 
respectively. The data for H1N1
cases of different states of USA is available on the website of
Centers for Disease Control and Prevention (CDC)
(http://www.cdc. gov/h1n1flu/), and the data of different provinces
of China is available from Sina.com
(http://news.sina. com.cn/z/zhuliugan/). The data for populations and
GDPs of different countries are obtained from English Wikipedia
(http://en.wikipedia.org/wiki/List\_of\_countries\_by\_population) and
(http://en.wikipedia. org/wiki/List\_of\_countries\_by\_GDP). There are
three different lists of GDPs and what we used here is the one from
the World Bank, which includes $182$ countries. Among the $135$
countries having reported the confirmed H1N1 cases until July 6,
$22$ of which do not have GDP data. They are all small
countries and the number of confirmed cases in these countries is
also quite few (the total of the $22$ countries are $163$ until July
6). We thus ignore them in evaluating the correlation in Fig. 3(c).


\subsection{Zipf's Law and Power Law}
Zipf's plot is widely used in the statistical analysis of the small-size sample \cite{Zipf}, which can be obtained by first rearranging
the data by decreasing order and then plotting the value of each data point versus its rank. The famous \emph{Zipf's law} describes a
scaling relation, $z(r)\sim r^{-\alpha}$, between the value of data point $z(r)$ and its rank $r$. As a signature of complex systems,
the Zipf's law is widely observed \cite{Newman2005,Clauset2009}. Indeed, it corresponds to a power-law probability density function $p(z)\sim z^{-\beta}$ with $\beta=1+\frac{1}{\alpha}$.

The Heaps' law \cite{Heaps}  is another well-known scaling law observed in many complex systems, which describes a sublinear
 growth of the number of distinct sets  $M (t) $ as the increasing
 of the total number of elements $N_T(t)$ belonging to those sets, with the power -law form $M \sim N_T^{\lambda}$.
Recent empirical analysis \cite{Zhang2008,Cattuto2009} suggested  that the Heaps' law  and
Zipf's law usually coexist. Actually, L\"u \emph{et al.} \cite{Lu2009} proved that if an evolving system has a stable
Zipf's exponent, its growth must obey the Heaps' law with exponent $\lambda\approx 1/\alpha$ (an approximate
estimation when $\alpha>1$: the larger the $\alpha$, the more accurate the estimation).

\subsection{Kendall's Tau}
In the empirical analysis, the numbers of confirmed cases,
populations and GDPs for different countries are very heterogeneous,
covering several orders of magnitude (e.g., the
population of China is about $2 \times 10^4$ times larger than
that of Dominica).  Thus the classical measurement like the Pearson
coefficient is not suitable in analyzing the correlations. We
therefore use the rank-based correlation coefficient named
\emph{Kendall's Tau}.
For two
series $\overrightarrow{x}=\{x_1,x_2,\cdots,x_m\}$ and
$\overrightarrow{y}=\{y_1,y_2,\cdots,y_m\}$, the Kendall's Tau is
defined as \cite{Kendall1938}
\begin{equation}
\tau_K = \frac{2}{m(m-1)}\sum_{i<j}\texttt{sgn}[(x_i-x_j)(y_i-y_j)],
\end{equation}
where $\texttt{sgn}(x)$ is the signum function,
which equals +1 if $x>0$, -1 if $x<0$, and 0 if $x=0$. $\tau_K$
ranges from +1 (exactly the same ordering of $\overrightarrow{x}$
and $\overrightarrow{y}$) to -1 (reverse ordering of
$\overrightarrow{x}$ and $\overrightarrow{y}$), and two uncorrelated
series have $\tau_K\approx 0$.

\subsection{On Power-Law Fitting}
Most of the distributions $P$ generated by simulations of our model with large
$\beta_1$ ($\geq 0.6$)  trend to a power-law-like type after
several steps of evolution. In the fittings of simulation results, we firstly judge if the curve of $P$
in this range is power-law-like. If yes, we fit the
curve by linear function in log-log plots in using least square fit
method to get the fitting parameters.
The range of the power-law fittings  is from $2$ to $50$. If there
is obvious deviation from power-law in this range, we do not use
power-law to fit the curve. The only exception is the distribution
$P$ when $b = 0.02$ in Fig. 8(a), where the range is from $1$ to
$30$, because the cut-off appears at rank $= 30$ due to slow spreading of the disease.
All the power-law fitting results in the model does not show the error-bar (e.g., the
dependence of $\alpha $ on various parameters of the model),
because the fitting error on the power-law exponent is far less than
the value of $\alpha$ for most cases after $10^4$ averages (e.g.,
$\alpha = 1.666 \pm 0.003$ when $\beta_1 = 0.8$, $\beta_2 = 0.2$,
$\rho =0.2$, $b = 0.06$ and $g = 0.2$ in the minimal model).

\subsection{Target Control}
When $s_i$ is highly heterogeneous, the nodes with the
largest $s_i$ will have the largest number of interregional travels in our model and have  high probability to spread
the disease. Target control on such nodes may efficiently reduce the spreading. To investigate the impact of the target control, we rank $s_i$ in the descending order,
and put the first $R_m$ nodes in the ranking series as the targets of strong border
control. In particular, we take  $\beta_1 > 0$ in Eq. (2) for the first $R_m$
nodes with the largest $s_i$  and $\beta_1 = 0$ for the  others.

\section*{Acknowledgments}





\newpage

\section*{Tables}

\begin{table}
\tabcolsep 0pt
\caption{The top 20 countries in the ranking and their total of confirmed cases in five typical dates} \vspace*{-10pt}
\small{}
\begin{center}
\def\temptablewidth{1.0\textwidth}
{\rule{\temptablewidth}{1pt}}
\begin{tabular*}{\temptablewidth}{@{\extracolsep{\fill}}ccccccc}
Rank & May 1 & May 10 & May 20 & June 10 & July 6 \\\hline
1 & Mexico 156 & U. S. A. 2254 & U. S. A. 5469 & U. S. A. 13217 & U. S. A. 	33902\\
2 & U. S. A. 141	& Mexico 1626 & Mexico 3648 & Mexico 5717 & Mexico 10262\\
3 & Canada 34 & Canada 280 & Canada 496 & Canada 2446 & Canada 7983\\
4 & Spain 13 & Spain 93 & Japan 210 & Chile 1694 & U. K. 7447\\
5 & U. K. 8	& U. K. 39 & Spain 107 &	Australia 1224 & Chile 7376\\
6&	Germany 	4	&	France 	12	&	U. K. 	102	&	U. K. 	666	&	Australia 	5298\\
7&	New Zealand 	4	&	Germany 	11		&Panama 	65	&	Japan 	485	&	Argentina 	2485\\
8&	Israel 	2	&	Italy 	9	&	France 	15	&	Spain 	331	&	China 	2101\\
9&	Austria 	1	&	Costa Rica 	8	&	Germany 	14	&	Argentina 	235	&	Thailand 	2076\\
10&	China 	1	&	Israel 	7	&	Colombia 	12	&	Panama 	221	&	Japan 	1790\\
11&	Denmark 	1	&	New Zealand 	7	&	Costa Rica 	9	&	China 	166	&	Philippines 	1709\\
12&	Netherlands 	1	&	Brazil 	6	&	Italy 	9	&	Costa Rica 	93	&	New Zealand 	1059\\
13&	Switzerland 	1	&	Japan 	4	&	New Zealand 	9	&	Dominican Rep.	91	&	Singapore	1055\\
14	&  & 			Korea, Rep. of 	3	&	Brazil 	8	&	Honduras	89	&	Peru 	916\\
15&	&			Netherlands 	3	&	China 	7	&	Germany 	78	&	Spain 	776\\
16	&	&			Panama 	3	&	Israel 	7	&	France 	71	&	Brazil 	737\\
17	&	&			El Salvador 	2	&	El Salvador 	6	&	El Salvador 	69	&	Israel 	681\\
18	&	&			Argentina 	1	&	Belgium 	5	&	Peru 	64	&	Germany 	505\\
19	&	&			Australia 	1	&	Chile 	5	&	Israel 	63	&	Panama 	417\\
20	&	&			Austria 	1	&	Cuba 	3	&	Ecuador 	60	&	Bolivia	416\\
    \end{tabular*}
       {\rule{\temptablewidth}{1pt}}
       \end{center}
\end{table}

\section*{Figure Legends}

\begin{figure}[!ht]
\begin{center}
 \centerline{\includegraphics[width=5.5in]{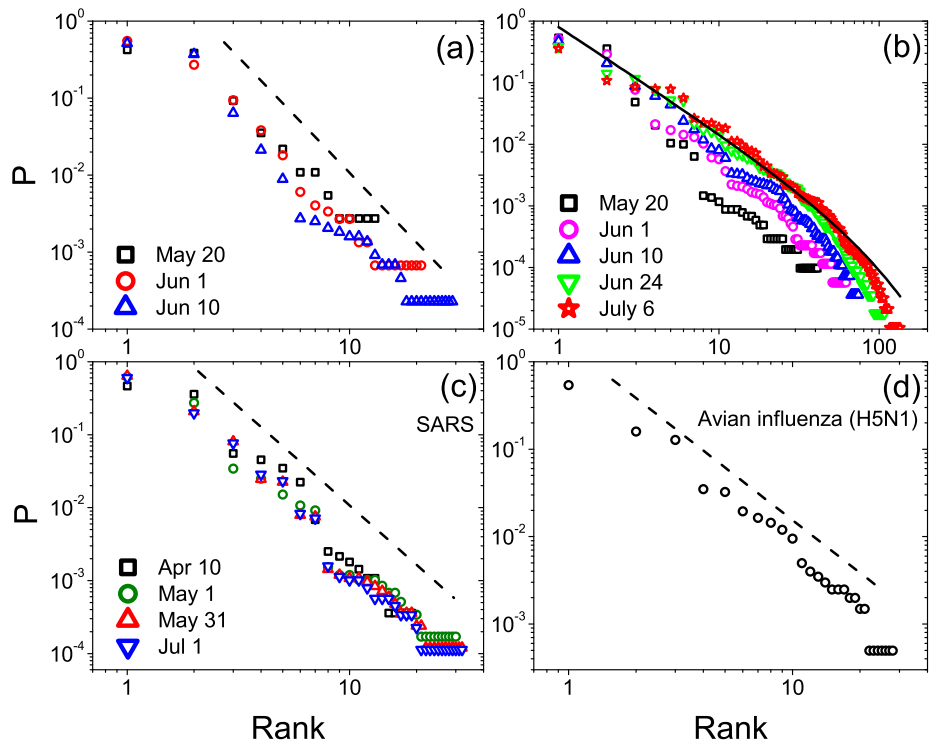}}
 \end{center}
  \caption{{\bf Zipf's distribution of various pandemic diseases}. (a) Zipf's distribution of the normalized  number of H1N1 cases in different countries in a log-log plot with date before May 15, 2009. (b) Same as (a) but with date after May 15, 2009. (c) Zipf's distribution of the normalized number of probable SARS cases for different countries in 2003. (d) Zipf's distribution of the normalized number of H5N1 cases for different countries in the whole year of 2008. The dash and solid curves are fitting functions described in the text.}
\end{figure}

\begin{figure}[!ht]
\begin{center}
\centerline{\includegraphics[width=5in]{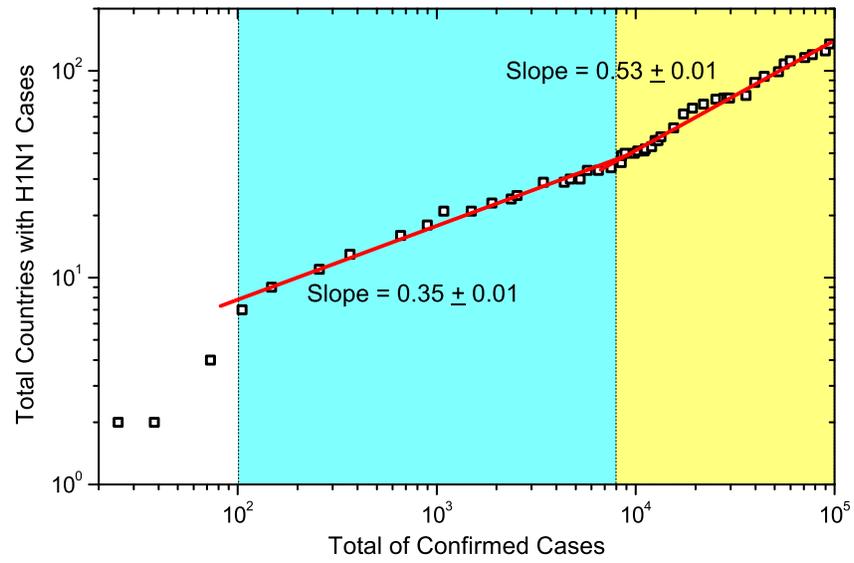}}
\end{center}
\caption{
{\bf Heaps' law in the spreading of H1N1}. Dependence between the number of infected countries, $M$,
and the global total $N_T$ of confirmed cases in log-log plot. The red and yellow areas indicate  power-law  fitting with  slope $0.35$ and $0.53$, respectively.
}
\end{figure}

\begin{figure}[!ht]
  \begin{center}
  \centerline{\includegraphics[width=5in]{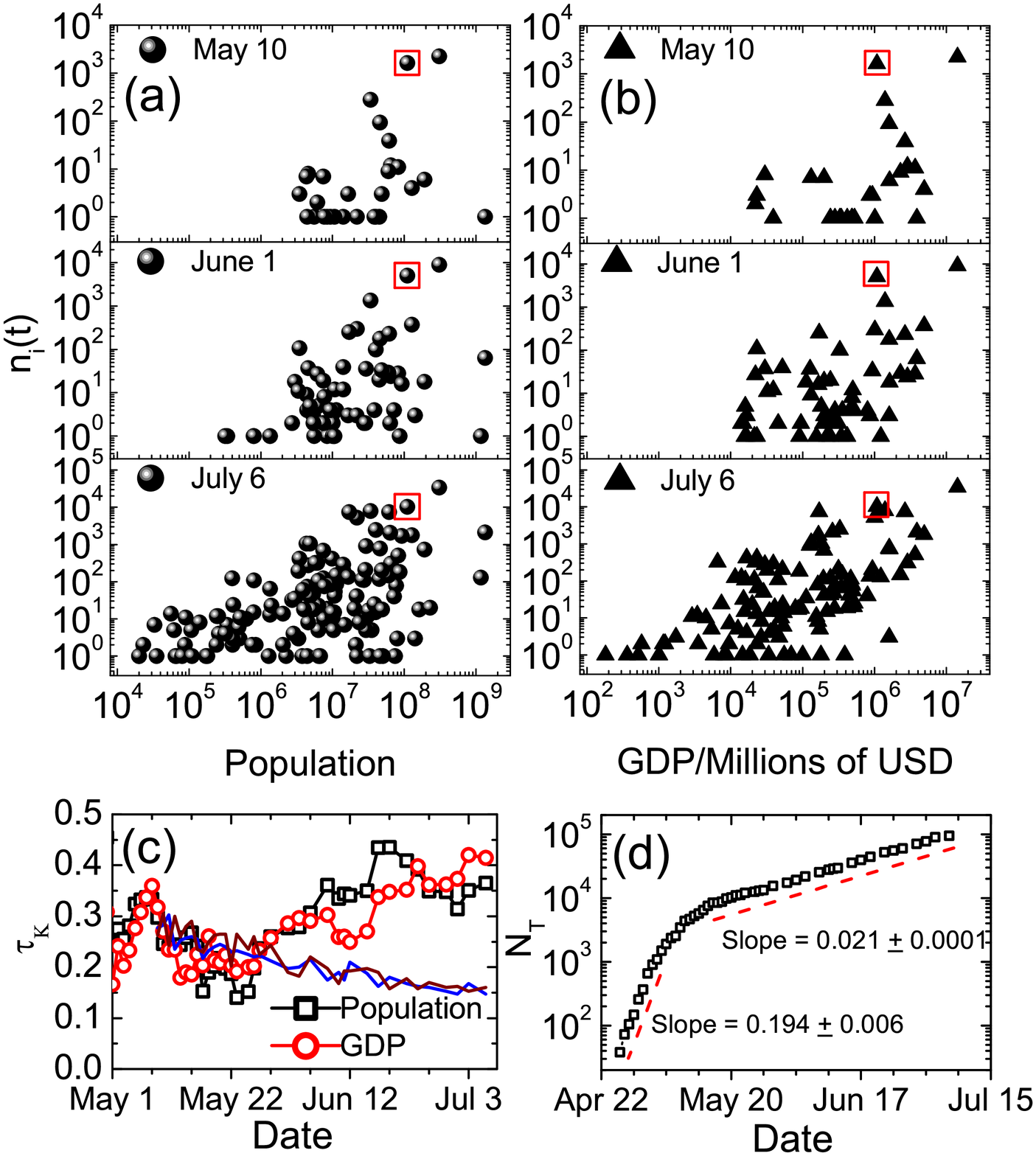}}
 \end{center}
  \caption{ {\bf Spreading of H1N1 depends on population and GDP.} (a) and (b) show the evolution of the dependence between the number of laboratory-confirmed cases $n_i$ for different countries and the population and GDP of these countries, respectively. The data for Mexico where the disease initiated are highlighted with an open square. (c) The Kendall's Tau $\tau_K$ of the correlations between the number of confirmed cases and the population/GDP. After about May 22, $\tau_K$ is clearly larger than the $95\%$ significance level (solid red and blue lines) of the surrogate data by randomly shuffling the order of the population/DGP of the countries having reported cases (see {\it Supporting information})}.
  (d) Growth of global total number $N_T$ of laboratory-confirmed cases of Influenza A in the semi-log plot.
\end{figure}

\begin{figure}[!ht]
  \begin{center}
  \centerline{\includegraphics[width=5in]{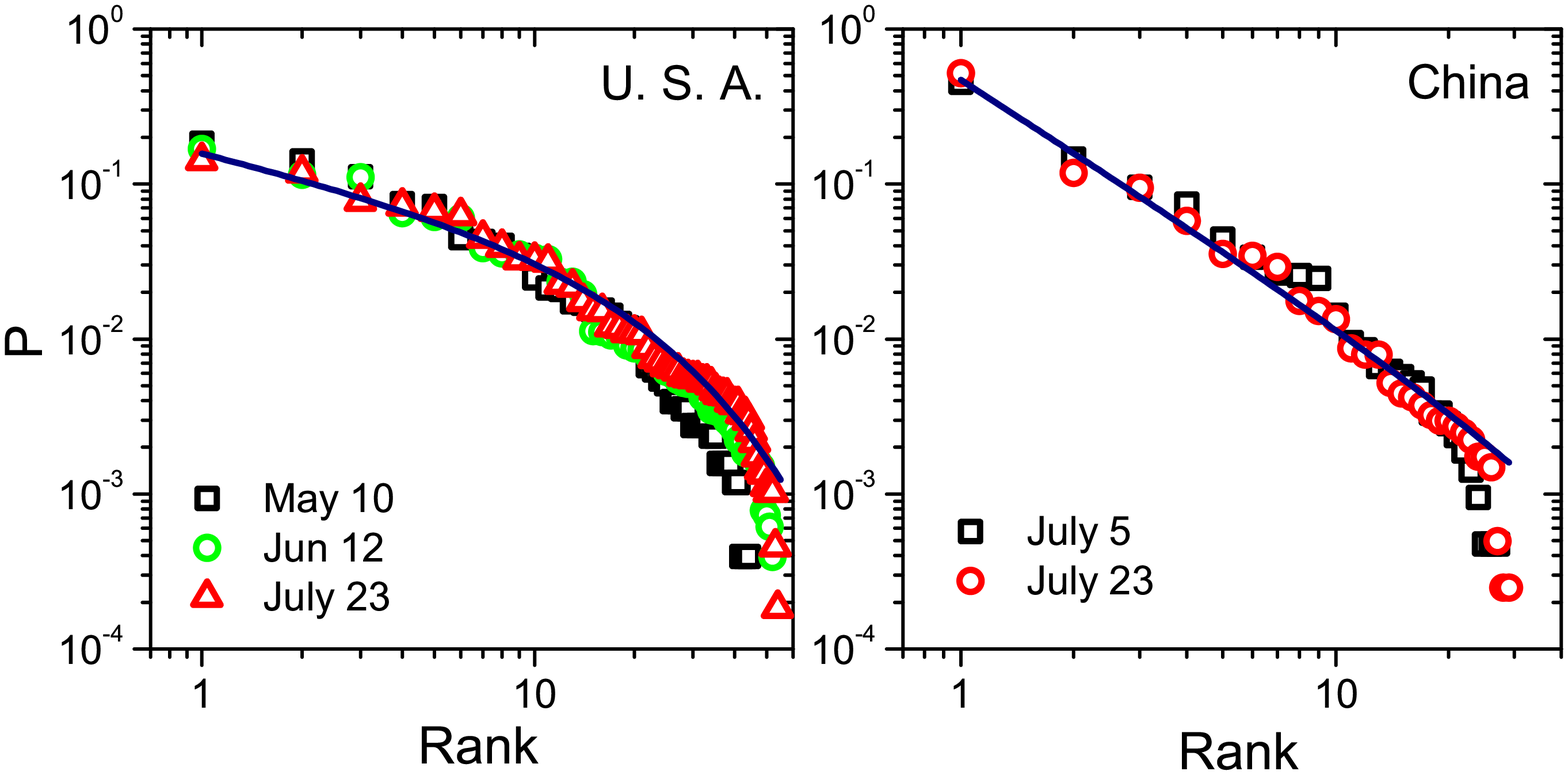}}
  \end{center}
  \caption{ {\bf Distribution of epidemic within countries.} Normalized distributions $P$ of confirmed cases in different states of USA and different provinces of China. For USA, the solid curve is $P = 0.17r^{-0.51}e^{-0.052r}$, fitting the data of July 23. For China, the solid line is of slope -1.79, fitting the data of July 23.}
\end{figure}

\begin{figure}[!ht]
\begin{center}
\centerline{\includegraphics[width=5in]{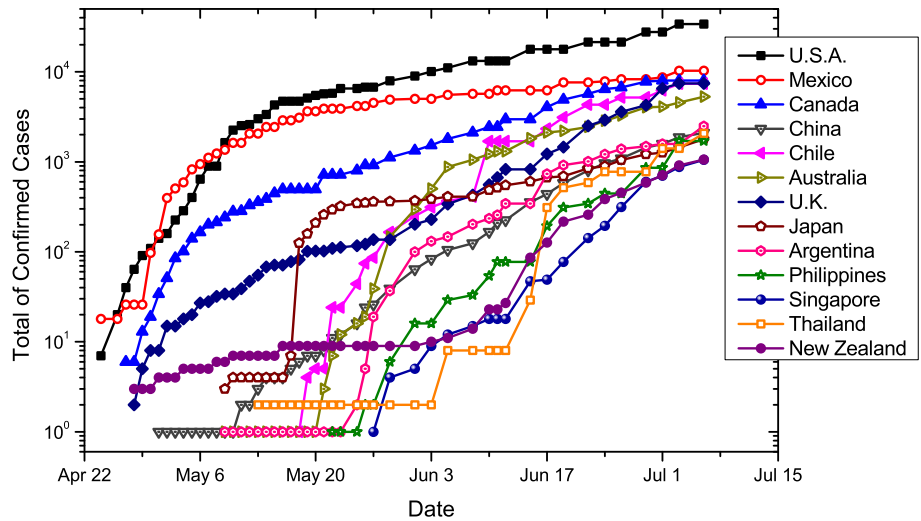}}
\end{center}
\caption{ {\bf The growth of  total confirmed cases within a country.} 
}
\end{figure}

\begin{figure}[!ht]
\begin{center}
 \centerline{\includegraphics[width=5in]{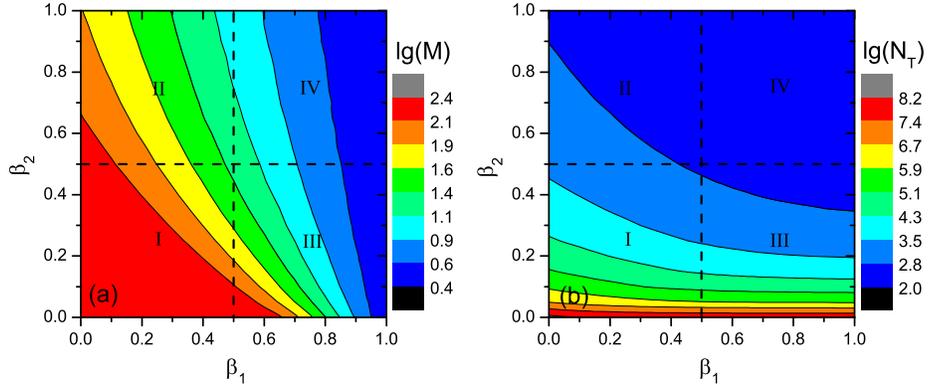}}
  \end{center}
  \caption{{\bf Effects of control on epidemic spreading.} Effects of the  parameters $\beta_1$ and $\beta_2$ on spreading speed, measured by the logarithm of the infected range  $M$   (a)  and the  logarithm of the total cases $N_T$ (b) at $t = 100$. Simulations run with  $b = 0.06$, and $g = 0.2$, and all the data are averaged over $10^4$ independent realizations.}
\end{figure}

\begin{figure}[!ht]
  \begin{center}
  \centerline{\includegraphics[width=5in]{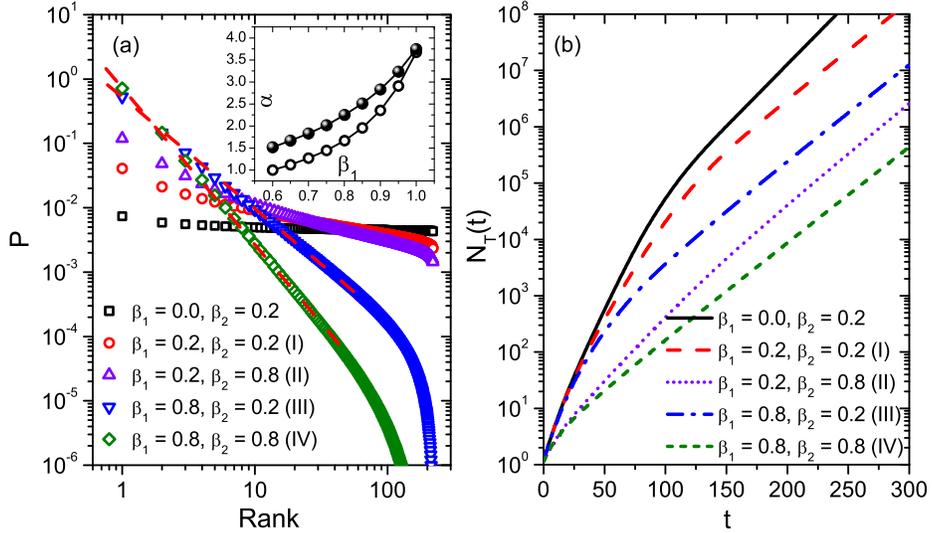}}
  \end{center}
  \caption{ {\bf Typical  patterns of spreading under various local and interregional control}.   (a) Typical normalized  distribution $P$ (at $t = 300$)  in the regime of I, II, III, and IV in the parameter space ($\beta_1, \beta_2$) shown in Fig. 5. The   two red dashed lines indicate the power-law functions with $\alpha=1.67$ and $\alpha=2.25$, respectively. The inserts show the dependence of $\alpha$ on $\beta_1$  for fixed $\beta_2 = 0.2$ (open circle) and $\beta_2=0.8$ (filled circle). (b)  The corresponding growth of $N_T$ with respect to time. The other parameters are the same as in Fig. 4.}
\end{figure}

\begin{figure}[!ht]
\begin{center}
\centerline{\includegraphics[width=5in]{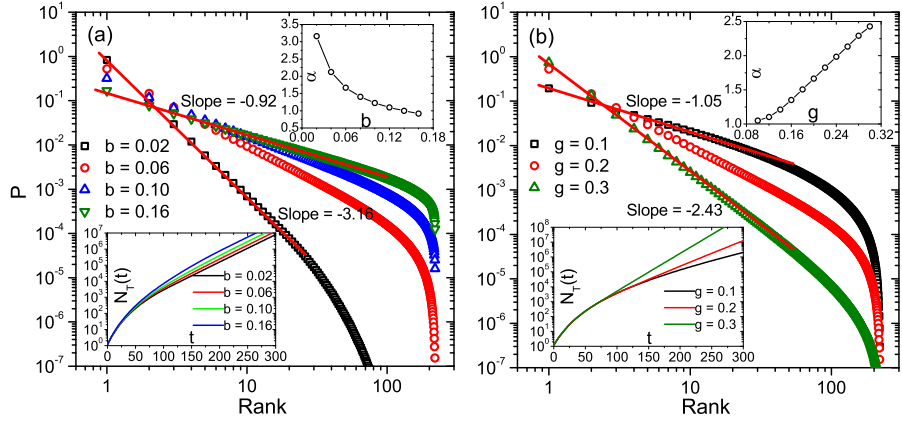}}
\end{center}
\caption{
{\bf Impacts of $b$ and $g$ on the normalized distributions $P$.} (a) The normalized distributions $P$ for different $b$.
The two red lines are power-law functions with exponent
$\alpha=0.92$ and $\alpha=3.16$, respectively. The top insert:
dependence of  $\alpha$ on $b$. Bottom insert: $N_T(t)$ vs. time for
various  $b$. Simulations run with $\beta_1 = 0.8$, $\beta_2 = 0.2$,
and $g = 0.2$. (b) The same as (a), but for different $g$.  The two
red lines are power-law functions with exponent $\alpha=1.05$ and
$\alpha=2.43$, respectively. Simulations run with $\beta_1 = 0.8$,
$\beta_2 = 0.2$, and $b = 0.06$. All the data are average over
$10^4$ independent runs.}
\end{figure}

\begin{figure}[!ht]
  \begin{center}
 \centerline{\includegraphics[width=5in]{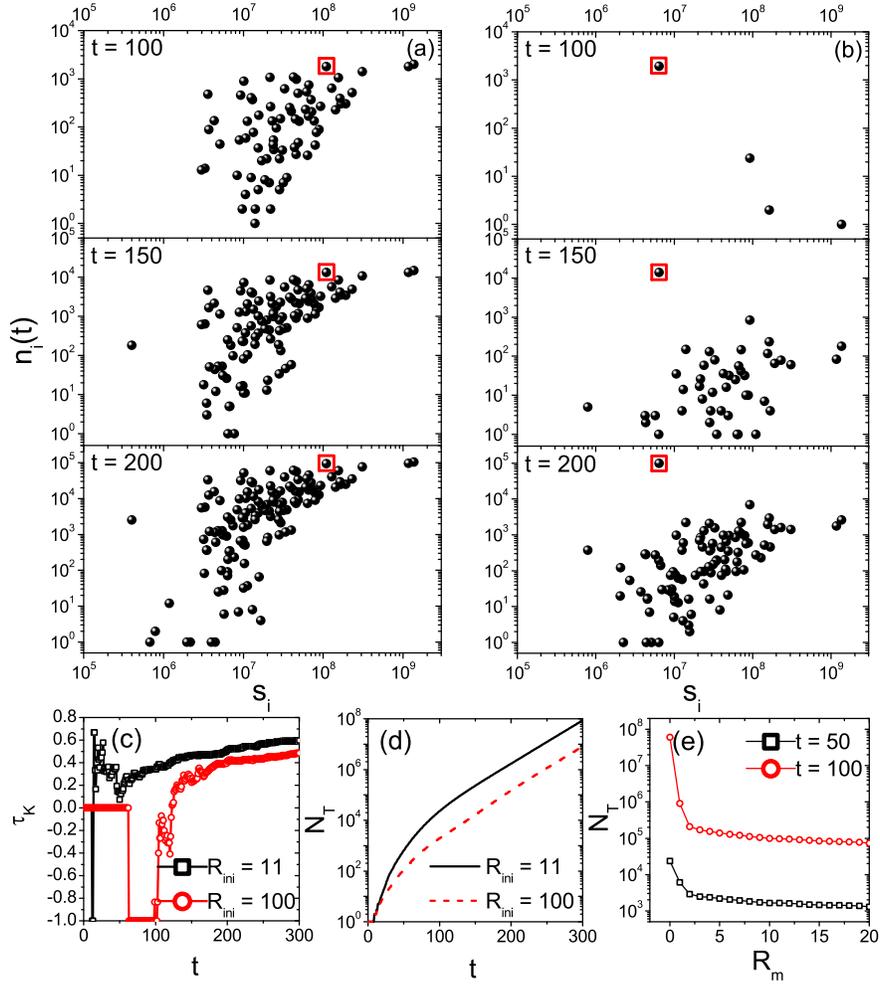}}
  \end{center}
  \caption{ {\bf Effects of heterogeneous $s_i$ in the model on epidemic spreading}.  Here $s_i$ is taken as the population of different countries.
 (a) and (b) show the evolution of  $n_i(t)$ vs $s_i$ when the disease is initiated at different countries
(highlighted by an open square). Simulations run with $\beta_1 = 0.8$,  $\beta_2= 0.2$, $b = 0.06$, and $g = 0.2$.
(c) Evolution of  $\tau_K$ between $n_i(t)$ and $s_i$.
(d) The corresponding growth of $N_T(t)$.  (a-d) are obtained from one realization of simulation.
Statistical results from many realizations are shown in Fig. S7 in SI.
(e) $N_T$ at $t = 50$ and $t = 100$ when $R_m$ nodes with the largest $s_i$ are the targets of control with $\beta_1=0.8$
(see {\sl Material and Methods}, the other parameters are the same as in (a-d)).
Data of (e) are averaged from $10^4$ independent runs with disease initiation at random nodes. }
\end{figure}

\begin{figure}[!ht]
\begin{center}
\centerline{\includegraphics[width=5in]{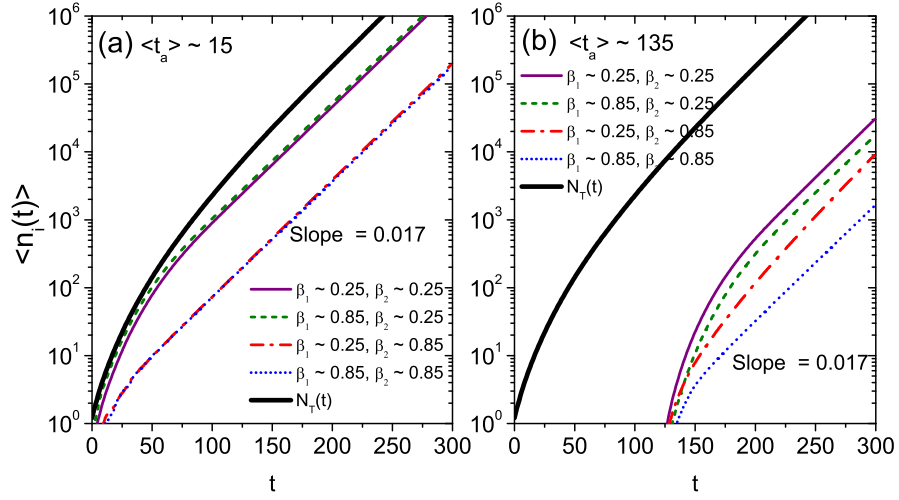}}
\end{center}
\caption{
{\bf The growth of averaged value of $n_i(t)$ for the nodes with different $\beta_2$, $g$ and the arrival time $t_a$.} (a) For the nodes that $0 \leq t_a < 20$ ($t_a \sim 15$). (b). For the nodes that $130 \leq t_a < 140$ ($t_a \sim 135$). In the two panels, $\beta_1 \sim 0.25$ denotes the data averaged for the nodes that $0.20 \leq \beta_1 < 0.30$, and $\beta_1 \sim 0.85$ for $0.80 \leq \beta_1 < 0.90$, and the same  for  $\beta_2$. Simulations run with  $\rho = 0.2$ and $b = 0.06$ and $g=0.2$. All the data are average over $10^4$ independent runs.
}

\end{figure}

\newpage
\mbox{}
\newpage
\newpage
\mbox{}
\newpage

\newpage
\mbox{}
\newpage
\newpage
\mbox{}
\newpage

\newpage
\mbox{}
\newpage
\newpage
\mbox{}
\newpage

\section*{Supporting Information}

\subsection*{The robustness of  Zipf's  distribution}

In the surveillance of H1N1 spreading process of each countries, many reasons can lead to the
deviation of the reported number of confirmed cases from  real number. The ranks of the real number could be different from those of the reported number for some countries.
Here we discuss the impact of this deviation on Zipf's plots, and our results show  that this impact is slight on
the power-law-like distribution of the reported number.

Let us denote the real total number of cases of the $i$th country as  $n_i^{'} = n_i + \varepsilon_i$,
where the positive value $\varepsilon_i$ represents the error or variation in surveillance of the country. It is reasonable to assume that
$\varepsilon_i$ could be related to $n_i$ or the population $M_i$ of a country.
We consider three types of assumptions on such variation as follows:
 Type A, the variation is correlated with the reported total of cases, namely $\varepsilon_i = \sigma n_i \eta$, where
  $\eta$ is a positive random number and obeys the right-part standard Gaussian distribution;
 Type B, the variation is correlated with the population of the country, so $\varepsilon_i = \sigma M_i (\frac{\sum_i n_i }{\sum_i M_i})\eta$,
 where $M_i$ denotes the population of the country;
 Type C, the variation is correlated with both the reported total of cases and the population of the country, and we assume
 $\varepsilon_i = \sigma n_i M_i (\frac{\sum_i n_i }{\sum_i n_i M_i})\eta$. The terms $\frac{\sum_i n_i }{\sum_i M_i}$ and $\frac{\sum_i n_i }{\sum_i n_i M_i}$
 are introduced for the purpose of normalization so that $\sigma$ value in the three cases are comparable.

The Zipf's distribution of $n_i^{'}$ for these three types of variations are compared to the original distribution  in Fig. S1 (a), (b), (c), respectively,  for various $\sigma$ values.
While the variation proportional to population (Type B) can result in some deviation of the distribution from the original one, the other two types
has no obvious impact on the distribution even though the average variation magnitude $\sigma$ is large.
This result indicate that the power-law-like distribution in Zipf's plot is robust under the variation in Type A and C.

\begin{center}
\onepic{0.35}{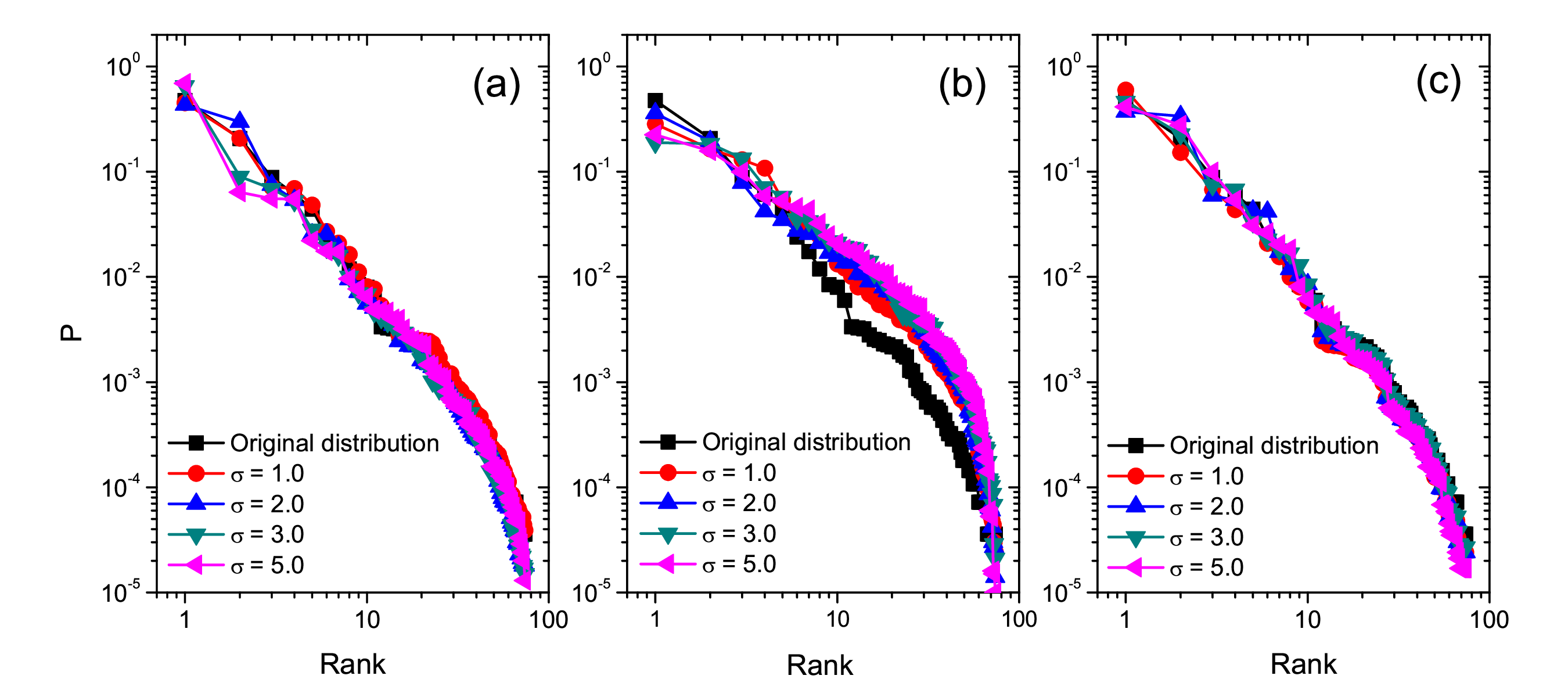}
\end{center}
{\small
Fig. S1. Comparison of the normalized Zipf's distribution of the three types of variation with  different $\sigma$ to  the original distribution of the total of confirmed cases
on a typical day (June 10, 2009). }

\vspace*{35pt}

\subsection*{Statistical testing of Kendall's Tau}

In this paper, the correlations between the total of confirmed cases of different countries and the populations or GDP of these countries are expressed by the value of Kendall's Tau $\tau_K$.
Here we test the significance of $\tau_K$ against the finite number of data points.

According to the algorithm of Kendall's Tau, $\tau_K \approx 0$ for two completely uncorrelated serials, however, a non-zero  value could be obtained due to the small number of countries  with reported cases,
 especially  in the early  stage of the spreading. To test the significance of the original $\tau_K$, we compare it to $\tau_K^{'}$  from surrogate data where the series of the population or GDP of the corresponding countries
  is randomly shuffled.  We can obtain a distribution of  $\tau_K^{'}$ for the surrogate for many realizations, and such a distribution is normal-like around $\tau_K^{'}=0$.
  From this distribution, we can obtain the $95\%$ significance levels.
 If the original $\tau_K$ is out of these levels, then there is less than  $5\%$ of possibility that the original  $\tau_K$ is due to coincidence in finite size uncorrelated series.
 Note that, in the early stage of May, there are just very few countries with reported cases.
 The number of shuffling realizations in the surrogate data  is too  small to obtain a reliable distribution of  $\tau_K^{'}$.

Fig. S2 shows the comparison between $\tau_K$ and the distribution of $\tau_K^{'}$.   While in the early stage (before   May 22)  we cannot reject the null hypothesis with very high confidence that the two series are uncorrelated,
$\tau_K$ is clearly significant afterwards for both the population and GDP.

\begin{center}
\onepic{0.45}{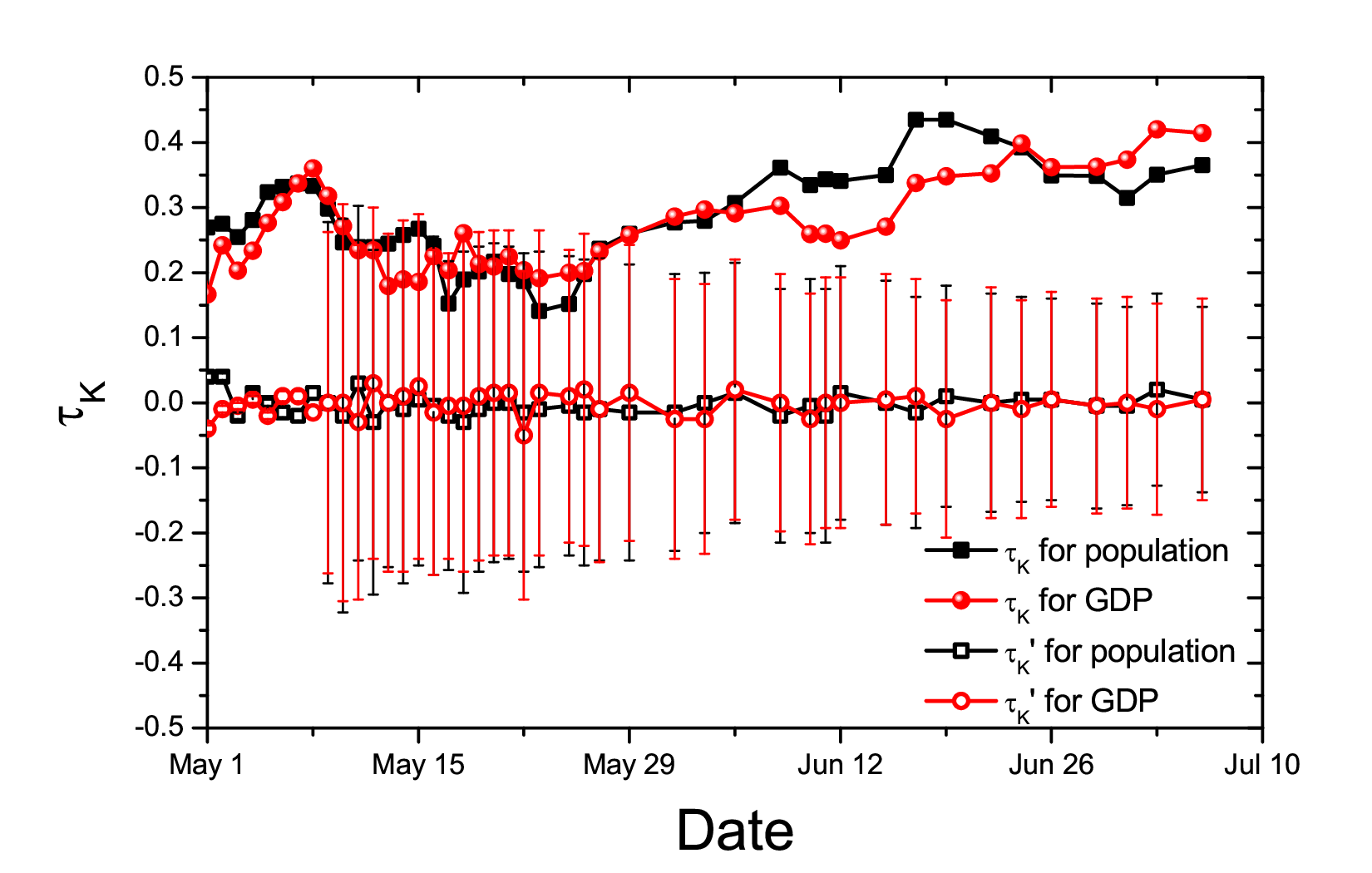}
\end{center}
Fig.  S2. Testing significance of Kendall's Tau that measures the correlations between total of confirmed cases and population / GDP of each of the countries. The range of error-bar  denotes $95\%$ of the significance level.  Since the number of countries with reported cases is few in the early stage of May,  the error-bars  are absent.
\vspace*{35pt}

\subsection*{Estimation of the reproductive number R in global level }

As shown in the main text (Fig. 3(d)), the growth of the global total of confirmed cases after the middle of May can be well fitted by a straight line with a slope 0.021 in semi-log plots, namely,
the growth follows a stable exponent form $\exp(\lambda t)$ with a daily  rate,  $\lambda = 0.021 \times \ln(10) = 0.048$. In our estimation of the reproductive number $R$,
we assumed that all the population is susceptible, and thus $R = R_0$, where $R_0$ is the basic reproductive number.
We used the formula in Ref. \cite{Li},  $R = 1 + V\lambda + f(1-f)(V\lambda)^2$ , where $f$ is the ratio of the infectious period to the serial interval, and $V$ is the mean serial interval, the sum of the mean infectious and mean latent periods. According to Ref. \cite{Cowl0, Cowl1}, the serial interval was estimated to be  in a range with mean 3.2
days and standard deviation 1.3 days. Therefore,  the range of the serial interval $V$ is set between 1.9 and 4.5 in our estimation. Assuming $f = 0.7$ or $0.3$ \cite{Li}, the reproductive number $R$ changes  from $1.09$ to $1.22$ when $V$  increases 1.9 days to 4.5 days (Insert  in Fig. S3). This range of $R$ value is slightly  less than the ranges obtained in several other researches based on the early period \cite{Fraser,Ni}, but in agreement with the results obtained from the data in the summer of 2009 \cite{Cowl1}.

We also investigated the evolution of the reproductive number $R$ in the pandemic Influenza A (H1N1) ( Fig. S3 (a)). The daily growth rate $\lambda$  are obtained from the growth of total of confirmed cases in each two or three days. The estimated value of $R$ is higher than 4 in the last a few days in April and then sharply reduces to a normal value between 1.0 and 1.4 after the middle of May. This reduction of $R$ may reflects the effect of various control and intervention schemes.

Interestingly, the evolution of the estimated reproduction number $R$ is related to the Zipf's distribution $P$. As seen in Fig. S3(b), during the period when  $R$ sharply reduces (before the middle of May), the exponent $\alpha$ of $P$ keeps in a higher level between $2.8$ and $3.7$. From the middle of May to July, $\alpha$ steadily declines from higher than $3$ to $1.7$ while  $R$ trends to stable.
$R$ and $\alpha$ appear to be positively correlated in the early stage of pandemic (insert in Fig. S3(b)).

\begin{center}
\onepic{0.38}{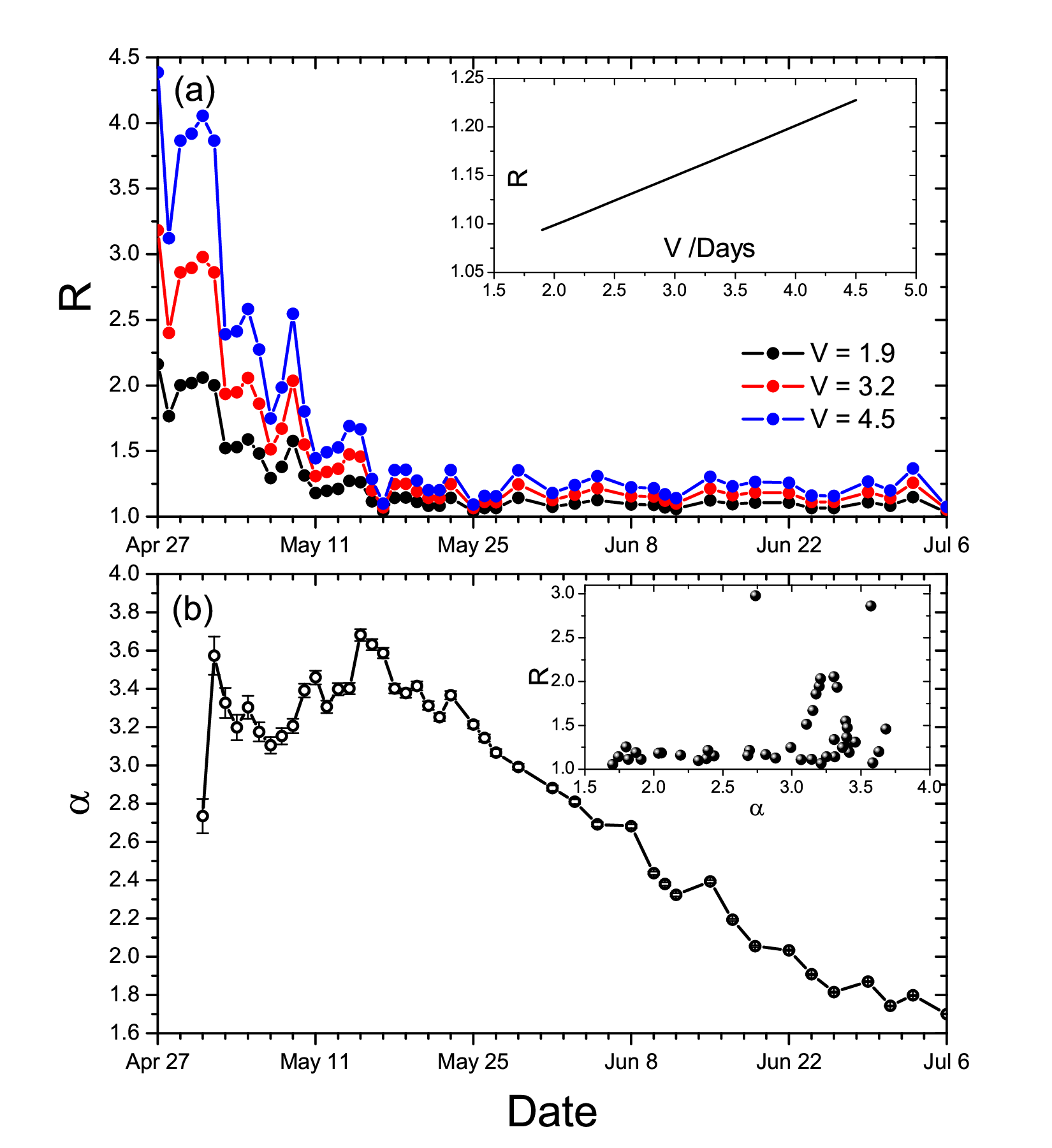}
\end{center}
{\small
Fig.  S3.  (a) Time evolution of the estimated reproduction number $R$ from April 28 to July 6. The insert shows  $R$  (of a given date) for different serial intervals $V$ from 1.9 days to 4.5 days. (b)  Time evolution of  the power-law exponent $\alpha$ of the normalized distribution $P$ of the total of confirmed cases, and the insert shows possible positive correlation $R$ and $\alpha$ when both of them are large in the early stage.}

\vspace*{35pt}

\subsection*{Evolution of the reproductive number R in our model }

The evolution of the reproductive number $R$ estimated from the growth of $N_T$ in our model is compared with $R$ estimated from real data. For the comparison,
 the time scale in our model is rescaled by the following method:
Firstly, we find two times $t_1$ and $t_2$ in the growth curve of $N_T$  in the model to satisfy $N_T(t_1) \approx 38$ and $N_T(t_2) \approx 94512$. The numbers $38$ and $94512$  are the global total of conformed cases in April 26 and July 6, respectively. There are 71 days form April 26 to July 6, thus we assume the  length of one time step in our model is corresponding to $\frac{71}{t_2 - t_1}$ days.
Consequently, the growth rate of each time step in our model is $\lambda(t) = [\ln(N_T(t)) - \ln(N_T(t))] (t_2 - t_1)/71$. From the growth rate $\lambda(t)$, the reproductive number $R(t)$ can be obtained as in the real data.
As shown in Fig. S4, in our typical parameter settings ($\beta_1 = 0.8$, $\beta_2 = 0.2$, $b = 0.06$ and $g = 0.2$), the evolution of $R$ also shows a rapid decrease: $R$ changes from the range between $1.7$ and $3.0$ to the range between $1.1$ and $1.4$, and generally in agreement with our empirical estimation of $R$.


\begin{center}
\onepic{0.38}{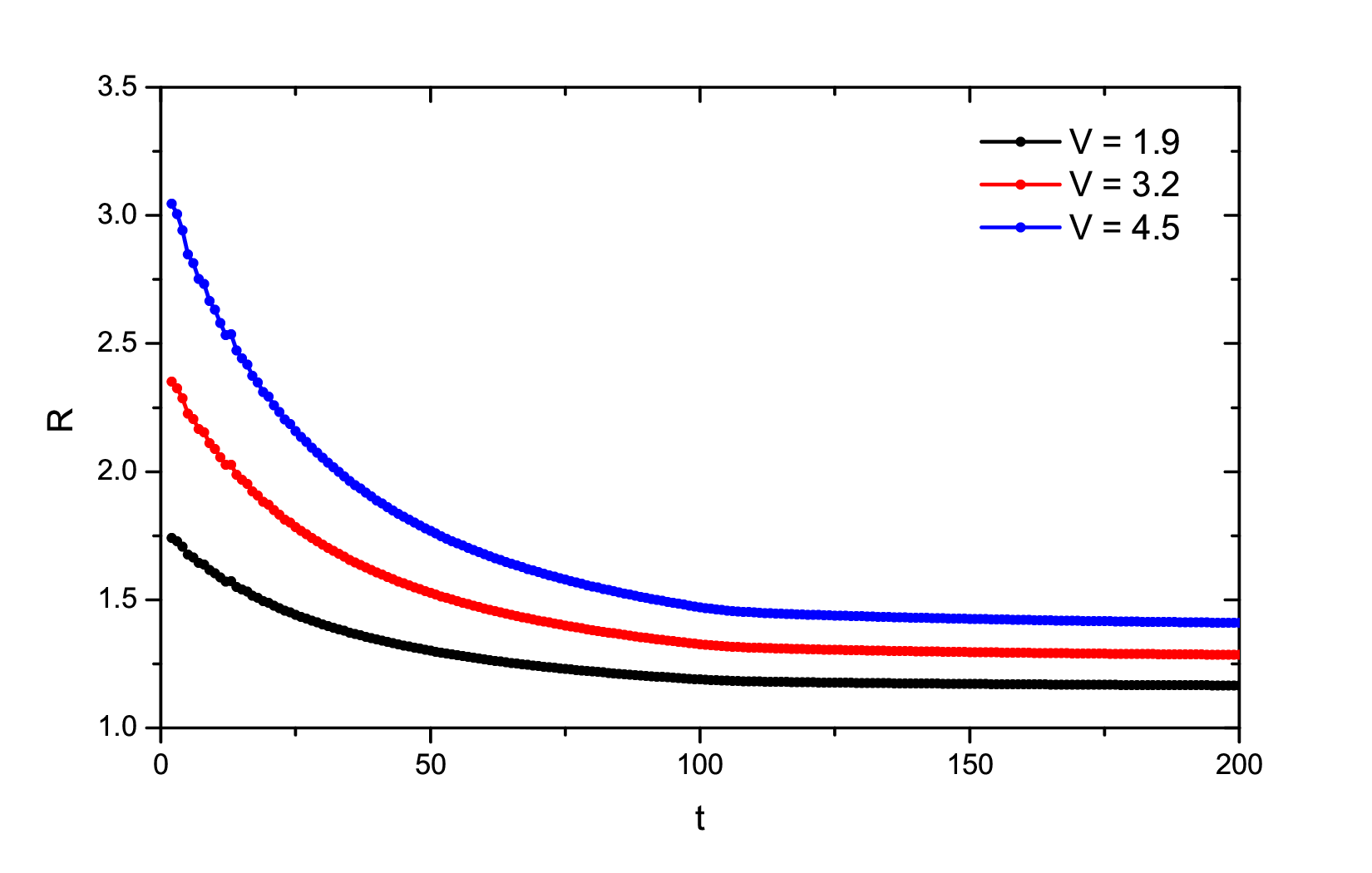}
\end{center}
{\small
Fig.  S4.  Time evolution of the  reproduction number $R$ estimated from the model  simulation result. Parameter settings are $\beta_1 = 0.8$, $\beta_2 = 0.2$, $b = 0.06$ and $g = 0.2$.}

\vspace*{35pt}

\subsection*{The evolution of  Zipf's  distribution  $P$ vs. the  Heaps'  plots  in the  model}

The empirical results in Fig. 1(b) in the paper indicates that the Zipf's plot  converges to a stable distribution before  the range of  spreading  $M$ reaches  saturation. The convergence to a stable distribution is inherent in our model. Let us consider a few nodes in the model with the largest $n_i$. The growth $n_i$ in such nodes is mainly determined by the local growth rate $\rho g$, $n_i(t+1) \approx  (1+\rho g) n_i(t)$, since the number of infected case due to input from other nodes is much smaller and can be neglected and the local growth has shifted to a stable rate due to local control in Eq. 4. When considering a power-law Zipf's distribution at time $t$,   $P_t (r)\sim r^{-\alpha}$,  the total global cases $N_T$ are  mainly contributed
by these  nodes with the largest $n_i$, i.e., $N_T(t+1) \approx (1+\rho g) N_T(t)$. Thus the normalized distribution  at $t+1$ for these nodes  is
 $P_{t+1}(r)= n_i(t+1)/N_T (t+1) \approx n_i(t)/N_T (t) \sim r^{-\alpha}$
which is invariant vs. time. This analysis is confirmed by the evolution of $P$ at various parameters  in Fig. S5(a-d). We can see that the distributions $P$ at different time overlap for the nodes with the smallest ranks $r$. The range of the forepart of the curve of $P$ which can be well fitted by power-law extends along with the time evolution, and the cut-off tail will move to large ranks $r$ till it reaches the system size $K$.

In our model,   the scaling property in the distribution $P$ is mainly contributed by large $\beta_1$. An extreme situation is that $\beta_1 > 0$ and $\beta_2 = 0$, namely, the effect of local control is ignored  (the parameter $g$ does not have  any impacts). In this case, $n_i(t+1)  \approx  (1+\rho) n_i(t)$ for the nodes with the largest $n_i$ and $N_T(t+1)\approx (1+\rho) N_T(t)$.  $P$ converges quickly to a power-law distribution  when $\beta_1$ is large (Fig. S5(a)), and the exponent $\alpha$ is quite large because $n_i$ of early infected nodes grow very fast, leading to a heterogeneous distribution. On the other hand when $\beta_2$ is large enough, the growth of $n_i$ and $N_T$ will shift quickly to a stable rate $\rho g$ and $P$ again converges to a power-law distribution.  $\alpha$ is significantly smaller than that at $\beta_2=0$ because the local control reduces significantly the growth rate of the early infected nodes, and $n_i$ is not as  heterogeneous. The stronger the control, the slower the growth of the newly infected nodes, and the more heterogeneous the distribution. Therefore, the  convergent exponent  $\alpha$ becomes  larger when $\beta_2$ increases. When $\beta_2$ is small, it takes a long period of time for infected nodes to achieve an stable exponential growth and consequently it takes  many steps for $P$ to converge.

In the discussions of the results of our model, the evolution time of the model generally is set as $300$ steps, because in the parameter settings in our discussion,  most of the $P$ distributions can show long range of power-law  part and the exponent of the power-law part  trends to stable
 after $300$ steps of  evolution.

From Fig. S5(a-d)  we can also  see that the distribution at a given time $t$ has a cut-off  at the range of spreading, i.e.  $r=M(t)$. At this point, we  have  $P(r)\approx 1/N_T(t)$ the last infect countries usually just has one or a few cases. When $\alpha$ is large (e.g., Fig. S5(a)), as an approximation we can assume that the power-law distribution $P(r) \sim r^{-\alpha}$ extends to the cut-off point, i.e., $P(r) \approx 1/N_T(t)\sim M^{-\alpha}(t)$; and we get $M \sim N^{\lambda}_T$ where $\lambda=1/\alpha$, implying that the Heaps' law \cite{Heaps} can be observed in the process (similar analysis could be found in \cite{A}, while for more accurate estimation, please see Ref. \cite{B}).  The  Heaps'  plots corresponding to the Zipf's plots in Fig. S5(a-d) are shown in Fig. S5(e-g). We can see that  the fitting exponents   satisfy  $\lambda \alpha \approx 1$ as expected from the analysis. The Heaps' plots also manifest  the saturation of $M$ when $N_T$ becomes very large.

\begin{center}
\onepic{0.28}{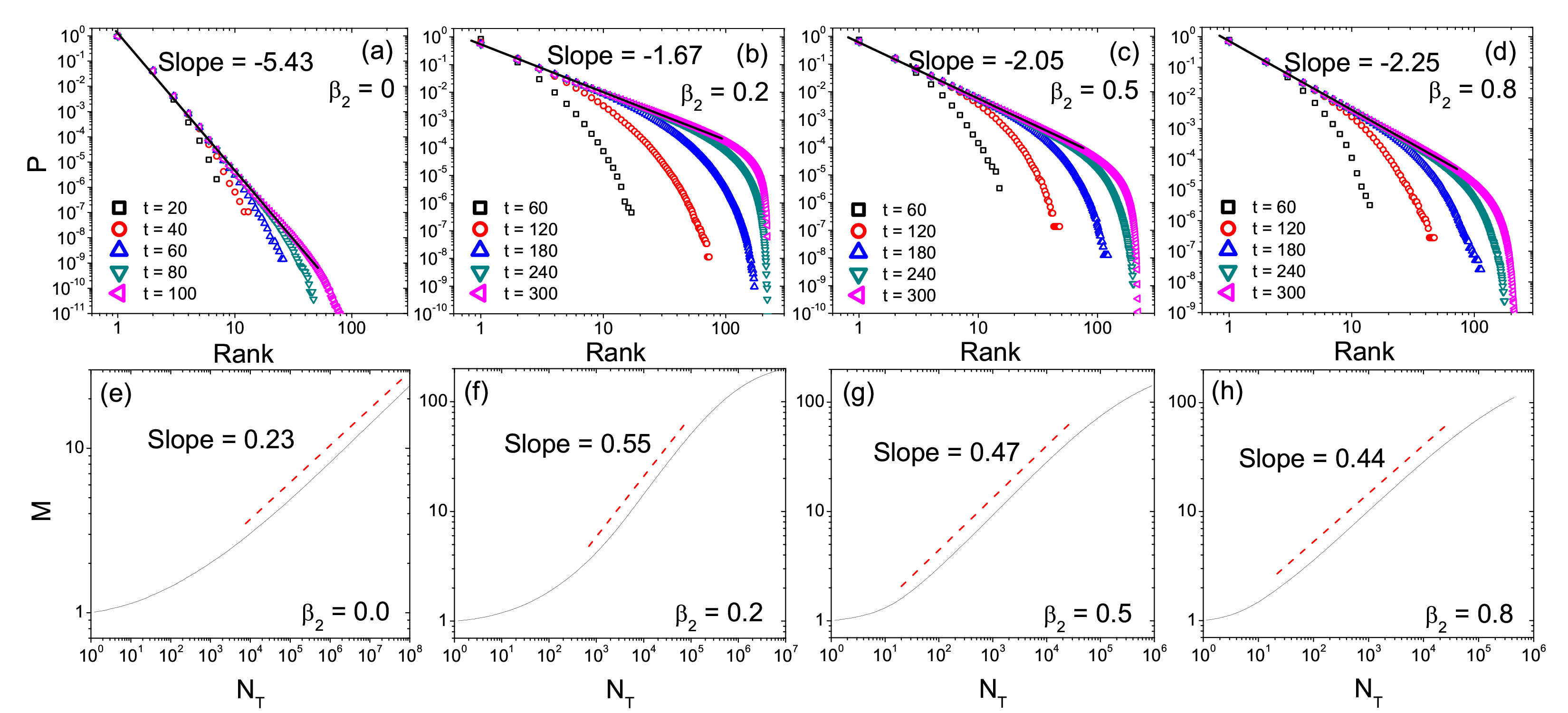}
\end{center}
{\small
Fig.  S5.  Upper panel (a-d): Evolution of the normalized Zipf's distributions $P$ at different time steps of the model for various $\beta_2$.
The other parameters are  $\beta_1 = 0.8$, $b = 0.06$, $\rho = 0.2$ and $g = 0.2$. All the data are average over $10^4$ independent runs.
Lower panel (e-h): Dependence between $M$ and $N_T$ (Heaps' plot)  generated by the model for different $\beta_2$,  corresponding  to the  distributions $P$ shown in (a)-(d), respectively.}

\vspace*{35pt}

\subsection*{Effects of heterogeneity in $s_i$}

In the paper we discuss the effect of heterogeneity in $s_i$ by taking $s_i$ as the population of a country. Fig.  9 in the paper  shows the evolution of the dependence between $n_i(t)$ and $s_i$  in one realization of the model simulation when the disease is initiated at two countries with population rank $R_{ini}=11$ and $R_{ini}=100$, respectively. Fig. S6 displays the statistics over many realizations:  (a) and (b) show the probability density function in the space ($\log_{10}s_i, \log_{10}n_i$) with color scale  and (c) and (d) are the time evolution of the total cases $N_T$ and the Kendall's  Tau, corresponding to Fig. 9 in the paper. In the early stage, the initiation position is the center of spreading, but shortly, the nodes with the largest $s_i$ become the super-spreaders. The spreading from the nodes with large $s_i$ to those with small $s_i$ becomes  very evident in this presentation.

\begin{center}
\onepic{0.45}{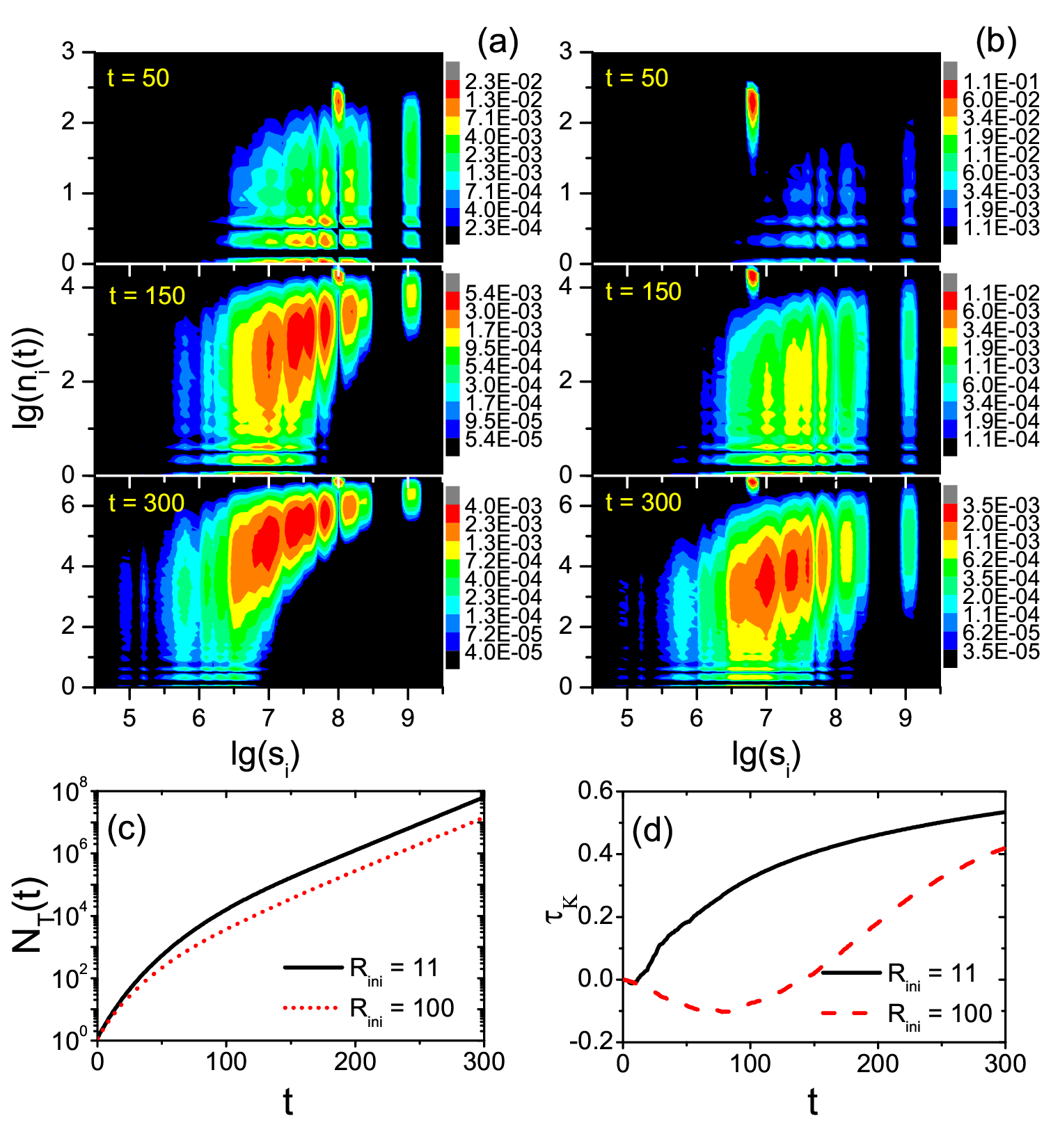}
\end{center}
{\small
Fig. S6.  Evolutions of the  probability density in the  space ($\log_{10}s_i, \log_{10}n_i$) obtained
from $10^4$ realizations of simulations with the epidemic initiation at node with population rank $R_{ini}=11$ (a) and $R_{ini}=100$ (b).
The corresponding growth of $N_T(t)$ (c) and evolution of Kendall's Tau $\tau_K$ between $n_i(t)$ and $s_i$  (d) averaged over all the realizations.
Simulations run on $\beta_1 = 0.8$,  $\rho = 0.2$, $b = 0.06$, and $g = 0.2$.
}
\vspace*{20pt}

In the following we carry out a more systematic analysis of the impact of heterogeneous $s_i$ and target control by considering power distributions
of $s_i$, i.e.,  $P(s) \sim s^{-\gamma}$.
 Two spreading processes are compared.  The first one is the situation without any control impacts, namely $\beta_1 = 0$ and $\beta_2 = 0$.
In the second situation we consider strong border control ($\beta_1 = 0.8$ and $\beta_2 = 0.2$,  typical parameter setting introduced in Fig. 7 in the paper).

Without control, the spreading is very fast even in the case of uniform $s_i$. Simulation results indicate that increased heterogeneity  at smaller $\gamma$ sharply accelerates the spreading by increasing the total cases $N_T$ in all the spreading period (Fig. S7(a)). The impact of heterogeneity  on the spreading range $M$ varies in different periods of the spreading: for stronger heterogeneity (smaller $\gamma$),  $M$ is larger  in  the early stage, but smaller  in the later stages (Fig. S7(b)). With control, the spreading is significantly suppressed (e.g., compare $t=100$ in panel (c) to $t=60$ in panel (a)), and the acceleration  by the heterogeneity is weaker: $N_T$ does not increase so strongly when $\gamma$ is smaller (Fig. S7(c)), while $M$ displays  similarly non-monotonic  but  relatively stronger dependence on $\gamma$ (Fig. S7(d)). The non-monotonic impact of heterogeneity on $M$ can  be understood as follows. When $s_i$ become rather heterogeneous,   the epidemic will rapidly arrive at the  nodes with large $s_i$ when initiated at a random node (see Fig. S7(b)), so $M$ is larger at smaller $\gamma$ in the early stage. Then the epidemic mainly grows in these a few early infected  nodes with the  largest $s_i$ and the majority of nodes with small $s_i$ have rather weak connections between them, which makes the spreading to new nodes more difficult, even though the total cases $N_T$ is larger. In the situations with control,  the spreading from these nodes having the largest $s_i$ and $n_i$  to the nodes with small $s_i$ is  further reduced. As a result, the  non-monotonic impact on $M$ is more obvious in the situations with control. The enhanced  spreading by the heterogeneity,  however, only makes the distribution $P$ slightly more homogeneous (Fig.  S8).

\begin{center}
\onepic{0.40}{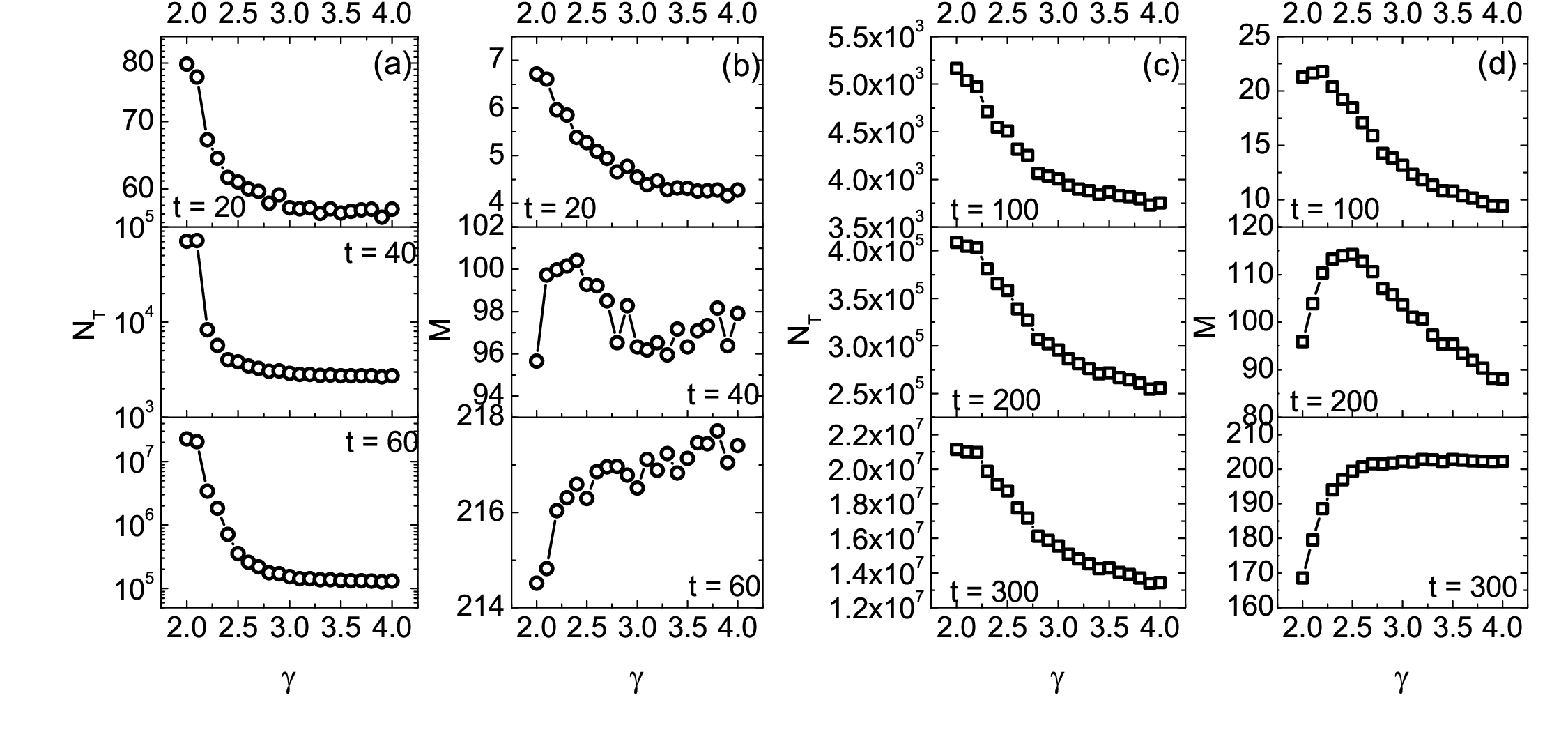}
\end{center}
{\small
Fig.  S7. Impacts of node heterogeneity ($\gamma$)
on the total cases $N_T$ and range $M$ of the epidemic spreading without control (panels (a) and (b): $\beta_1 = 0$ and $\beta_2 =0$) and
with control (panels (c) and (d): $\beta_1 = 0.8$ and $\beta_2 = 0.2$).  Note the different scales in the y-axes of  (a) and (c).
The simulations run on $\rho = 0.2$, $b = 0.06$, and $g = 0.2$. All of data are averaged from $10^4$ independent runs.
}
\vspace*{20pt}

The comparison of these heterogeneous networks with the minimal models show that while strong heterogeneity in the nodes (countries) could be an accelerating factor, just like the effect of the heterogeneous degree distribution of  complex networks \cite{Pastor0}, the strengths of control play a leading and dominant role in determining the epidemic spreading patterns. In fact, the impact of strong heterogeneity can be compensated with slightly increased border control parameter $\beta_1$.

\begin{center}
\onepic{0.35}{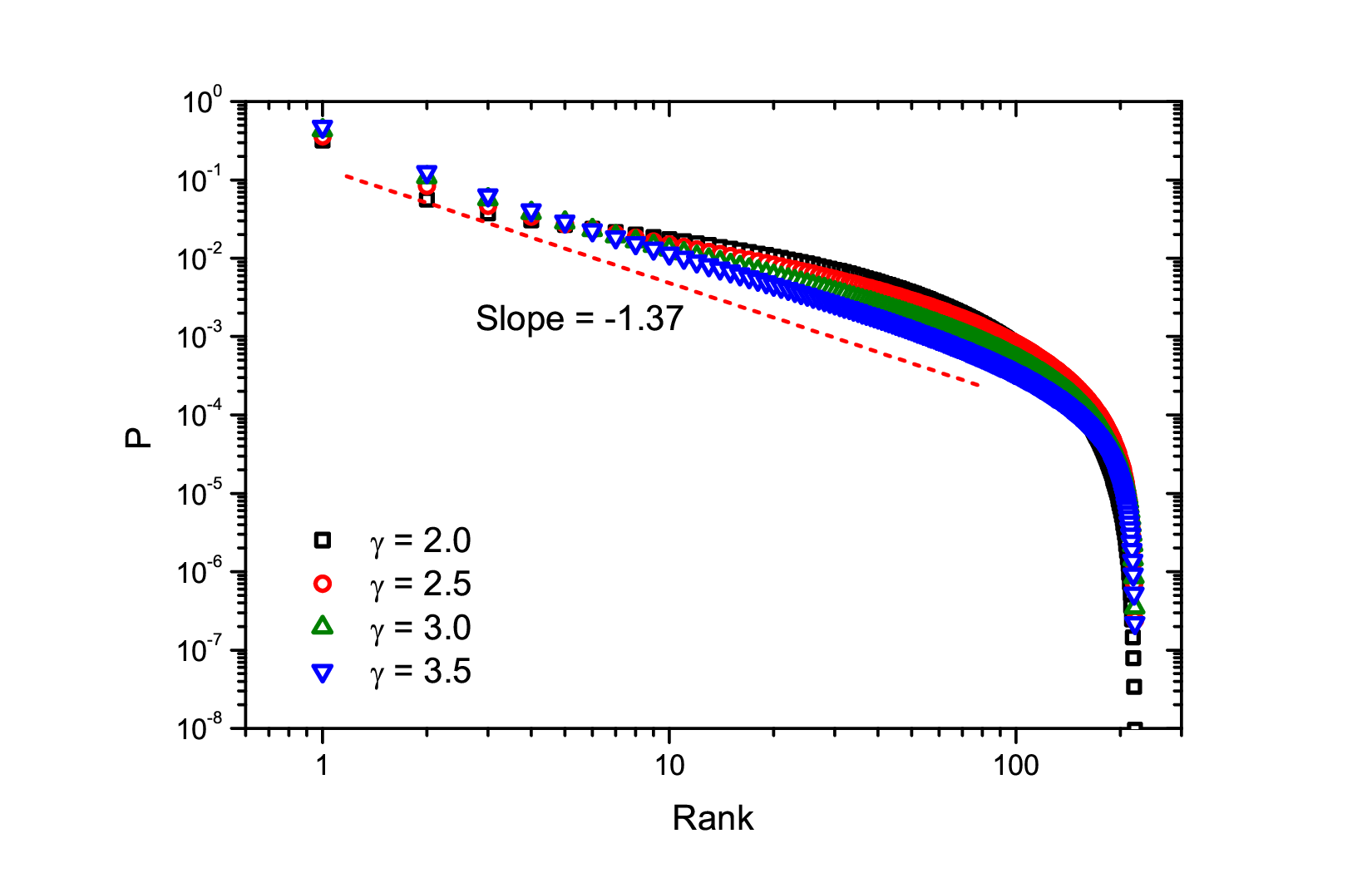}
\end{center}
{\small
Fig. S8. The distributions $P$ for different $\gamma$ at $t=300$. Simulation runs on $\beta_1 = 0.8$, $\beta_2 = 0.2$, $\rho = 0.2$, $b = 0.06$, and $g = 0.2$. The red dashed line denotes the power-law with exponent $-1.37$.  All of data are obtained at $t=300$ and averaged from $10^4$ independent runs.
}
\vspace*{20pt}

\begin{center}
\onepic{0.25}{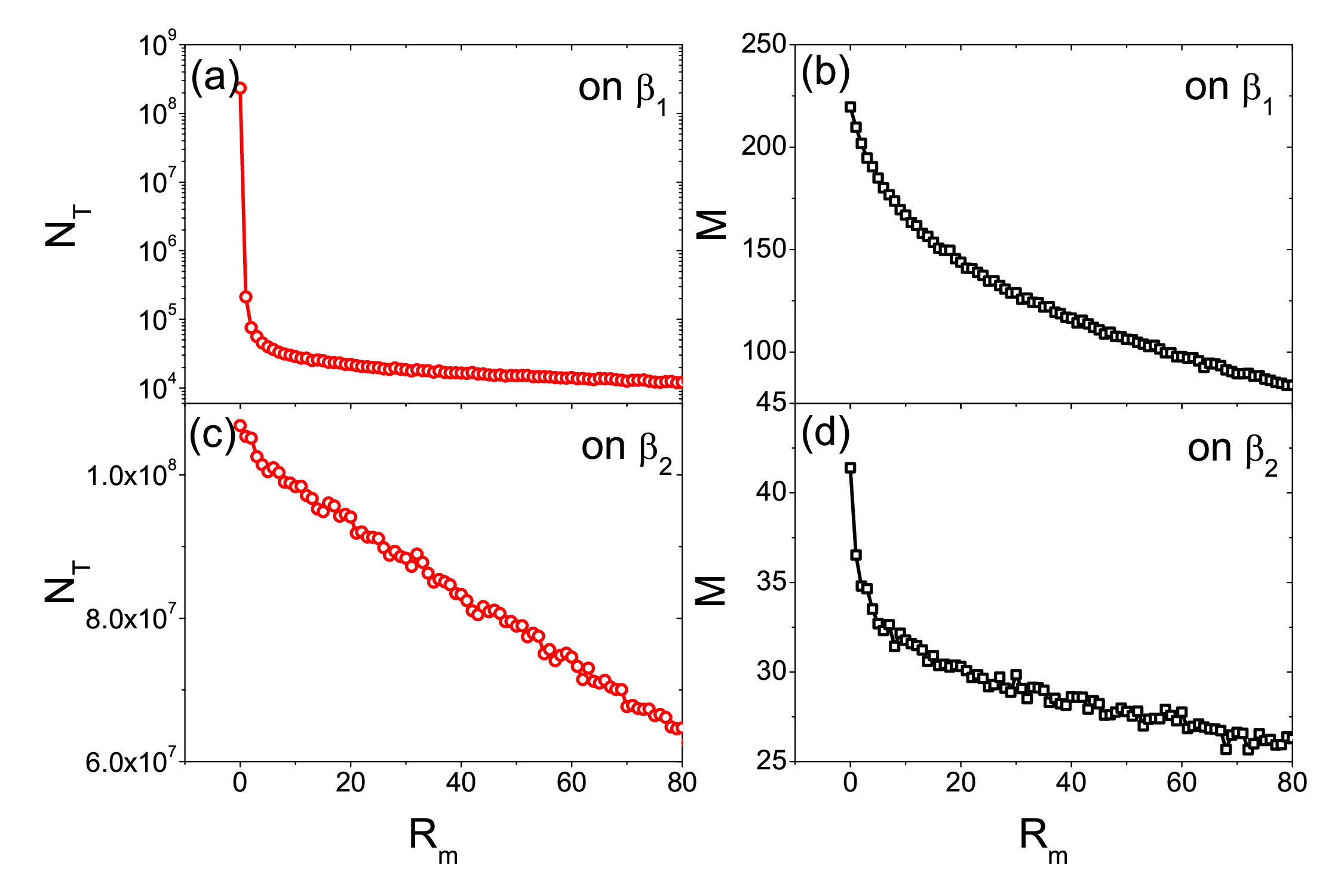}
\end{center}
{\small
Fig. S9. (a) and (b): Total  cases $N_T$ and range $M$ vs. $R_m$, the number of nodes with target border  control ($\beta_1 = 0.8$ for the  $R_m$ nodes with the largest $s_i$, and $\beta_1 = 0$ for others). Simulations run on $\gamma = 2.0$, $\beta_2 = 0.2$, $\rho = 0.2$, $b = 0.06$, and $g = 0.2$.   (c) and (d): as in (a) and (b), but with target local control ($\beta_2 = 0.2$ for the  $R_m$ nodes with the largest $s_i$, and $\beta_2 = 0$ for others).   Simulations run on $\gamma = 2.0$, $\beta_1 = 0.8$, $\rho = 0.2$, $b = 0.06$. All of data are obtained at $t=100$ and  averaged from $10^4$ independent runs.
}
\vspace*{20pt}

Generally speaking, the nodes with large intensity on heterogeneous structures usually are the  keys towards the dynamics of the system. We have applied the target control to the first $R_m$  nodes with the largest $s_i$. The spreading can be sharply decelerated by reducing both the total cases $N_T$ and range $M$ (Fig.  S9(a) and (b)), when only a few nodes with the largest  $s_i$ are in strong border control (setting  $\beta_1 > 0$ for the first $R_m$ nodes and $\beta_1 = 0$ for the  others). Similar impact  can also be  observed with  target  local control on the same nodes  (setting  $\beta_2 > 0$ for the front $R_m$ nodes and $\beta_2 = 0$ for others, see Fig.  S9(c) and (d)). In this case, infection in those nodes with $\beta_2=0$ grows very fast  (with a rate $\rho$) and the control on a few nodes does not reduce the total $N_T$ very significantly. However, $M$ is clearly reduced because the nodes with the largest $s_i$ are usually the centers of spreading in the early stage and the target control within these nodes will retard  the spreading to other nodes.   These impacts of control on the nodes with  large $s_i$  are quite similar to the targeted immunization strategy on hubs notes with the largest degrees in scale-free networks \cite{C}. Therefore, the heterogeneity could be employed to prevent the spreading with target  control strategies,  and the target control over a few nodes can save the overall cost of control.

\vspace*{35pt}


\begin{table}
\tabcolsep 0pt
\caption{Rank of total of confirmed cases for different countries (I)} \vspace*{-10pt}
\tiny
\begin{center}
\def\temptablewidth{1.0\textwidth}
{\rule{\temptablewidth}{1pt}}
\begin{tabular*}{\temptablewidth}{@{\extracolsep{\fill}}ccccccc}
Rank & May 1 & May 10 & May 20 & June 10 & July 6 \\\hline
1 & Mexico 156 & U. S. A. 2254 & U. S. A. 5469 & U. S. A. 13217 & U. S. A. 	33902\\
2 & U. S. A. 141	& Mexico 1626 & Mexico 3648 & Mexico 5717 & Mexico 10262\\
3 & Canada 34 & Canada 280 & Canada 496 & Canada 2446 & Canada 7983\\
4 & Spain 13 & Spain 93 & Japan 210 & Chile 1694 & U. K. 7447\\
5 & U. K. 8	& U. K. 39 & Spain 107 &	Australia 1224 & Chile 7376\\
6&	Germany 	4	&	France 	12	&	U. K. 	102	&	U. K. 	666	&	Australia 	5298\\
7&	New Zealand 	4	&	Germany 	11		&Panama 	65	&	Japan 	485	&	Argentina 	2485\\
8&	Israel 	2	&	Italy 	9	&	France 	15	&	Spain 	331	&	China 	2101\\
9&	Austria 	1	&	Costa Rica 	8	&	Germany 	14	&	Argentina 	235	&	Thailand 	2076\\
10&	China 	1	&	Israel 	7	&	Colombia 	12	&	Panama 	221	&	Japan 	1790\\
11&	Denmark 	1	&	New Zealand 	7	&	Costa Rica 	9	&	China 	166	&	Philippines 	1709\\
12&	Netherlands 	1	&	Brazil 	6	&	Italy 	9	&	Costa Rica 	93	&	New Zealand 	1059\\
13&	Switzerland 	1	&	Japan 	4	&	New Zealand 	9	&	Dominican Rep.	91	&	Singapore	1055\\
14	&  & 			Korea, Rep. of 	3	&	Brazil 	8	&	Honduras	89	&	Peru 	916\\
15&	&			Netherlands 	3	&	China 	7	&	Germany 	78	&	Spain 	776\\
16	&	&			Panama 	3	&	Israel 	7	&	France 	71	&	Brazil 	737\\
17	&	&			El Salvador 	2	&	El Salvador 	6	&	El Salvador 	69	&	Israel 	681\\
18	&	&			Argentina 	1	&	Belgium 	5	&	Peru 	64	&	Germany 	505\\
19	&	&			Australia 	1	&	Chile 	5	&	Israel 	63	&	Panama 	417\\
20	&	&			Austria 	1	&	Cuba 	3	&	Ecuador 	60	&	Bolivia	416\\
21	&	&			China 	1	&	Guatemala 	3		& Guatemala 	60	&	Nicaragua	321\\
22	&	&			Colombia 	1	&	Korea, Rep. of 	3	&	Philippines 	54	&	El Salvador 	319\\
23	&	&			Denmark 	1	&	Netherlands 	3	&	Italy 	50	&	France 	310\\
24	&	&			Guatemala 	1	&	Peru 	3		&Korea, Rep. of 	48	&	Guatemala 	286\\
25	&	&			Ireland	1	&	Sweden 	3	&	Brazil 	36	&	Costa Rica 	277\\
26	&	&			Poland 	1	&	Finland 	2	&	Colombia 	35	&	Venezuela	206\\
27	&	&			Portugal 	1	&	Malaysia 	2	&	Nicaragua	29	&	Ecuador 	204\\
28	&	&			Sweden 	1	&	Norway 	2	&	Uruguay	24	&	Korea, Rep. of 	202\\
29	&	&			Switzerland 	1	&	Poland 	2	&	New Zealand 	23	&	Uruguay	195\\
30	&  &  &						Thailand 	2	&	Netherlands 	22	&	Viet Nam 	181\\
31	&  &  &						Turkey 	2	&	Kuwait	18	&	Greece	151\\
32	&  &  &							Argentina 	1		& Singapore	18	&	Italy 	146\\
33	&  &  &							Australia 	1	&	Paraguay	16	&	Netherlands 	135\\
34	&  &  &							Austria 	1	&	Sweden 	16	&	India 	129\\
35	&  &  &						Denmark 	1	&	Switzerland 	16	&	Brunei Darussalam	124\\
36	&  &  &							Ecuador 	1	&	Viet Nam 	15	&	Honduras	123\\
37	&  &  &							Greece	1	&	Belgium 	14	&	Colombia 	118\\
38	&  &  &							India 	1	&	Ireland	12	&	Saudi Arabia	114\\
39	&  &  &							Ireland	1	&	Venezuela	12	&	Malaysia 	112\\
40	&  &  &							Portugal 	1		&Turkey 	10	&	Cyprus	109\\
41	&  &  &							Switzerland 	1	&	Norway 	9		&Dominican Rep.	108\\
42		&  &  &  &								Romania	9	&	Paraguay	106\\
43	&  &  &  &									Denmark 	8	&	Cuba 	85\\
44	&  &  &  &									Egypt	8	&	Sweden 	84\\
45	&  &  &  &									Lebanon	8	&	Egypt	78\\
46	&  &  &  &									Thailand 	8	&	Switzerland 	76\\
47	&  &  &  &									Jamaica	7	&	Ireland	74\\
48		&  &  &  &								Poland 	6	&	Denmark 	66\\
49	&  &  &  &									Austria 	5	&	Trinidad and Tobago	65\\
50	&  &  &  &									Cuba 	5		& West Bank and Gaza Strip 	60\\
51	&  &  &  &									Greece	5	&	Belgium 	54\\
52	&  &  &  &									Malaysia 	5	&	Lebanon	49\\
53	&  &  &  &									Estonia	4	&	Finland 	47\\
54	&  &  &  &									Finland 	4	&	Portugal 	42\\
55	&  &  &  &									India 	4	&	Norway 	41\\
56	&  &  &  &									Bolivia	3	&	Romania	41\\
57	&  &  &  &									Hungary	3	&	Turkey 	40\\
58	&  &  &  &									Russia 	3	&	Kuwait	35\\
59	&  &  &  &									Slovakia	3	&	Jamaica	32\\
60	&  &  &  &									Bahamas	2	&	Poland 	25\\
61	&  &  &  &									Barbados	2	&	Malta	24\\
62	&  &  &  &									Bulgaria	2	&	Jordan	23\\
63	&  &  &  &									Czech Republic	2	&	Qatar	23\\
64	&  &  &  &									Iceland	2	&	Indonesia	20\\
65	&  &  &  &									Portugal 	2	&	Austria 	19\\
66	&  &  &  &									Trinidad and Tobago	2	&	Sri Lanka	19\\
67	&  &  &  &									Bahrain	1	&	Bangladesh	18\\
68	&  &  &  &									Cayman Islands, UKOT	1	&	Slovakia	18\\
69	&  &  &  &									Cyprus	1	&	South Africa	18\\
70	&  &  &  &									Dominica	1	&	USA Puerto Rico	18\\
71	&  &  &  &									Luxembourg	1	&	Morocco	17\\
72	&  &  &  &									Saudi Arabia	1	&	Bahrain	15\\
73	&  &  &  &									Ukraine	1	&	Czech Republic	15\\
74	&  &  &  &									United Arab Emirates	1	&	Kenya	15\\
75		& & & & &											Serbia	15\\
76	& & & & &													Cayman Islands, UKOT	14\\
77	& & & & &													Slovenia	14\\
78	& & & & &													Estonia	13\\
79	& & & & &													Barbados	12\\
80	& & & & &													France, New Caledonia, FOC	12\\
81	& & & & &													Iraq	12\\
82	& & & & &													Hungary	11\\
83	& & & & &													Suriname	11\\
84	& & & & &													U. K., Jersey, Crown Dependency	11\\
85	& & & & &													Bulgaria	10\\
86	& & & & &													Montenegro	10\\
87	& & & & &													Netherlands Antilles, Curaçao *	8\\
88	& & & & &													United Arab Emirates	8\\
89	& & & & &													Yemen	8\\
90	& & & & &													Bahamas	7\\
    \end{tabular*}
       {\rule{\temptablewidth}{1pt}}
       \end{center}
\end{table}

\begin{table}[!h]
\tabcolsep 0pt \caption{Rank of total of confirmed cases for different countries (II)} \vspace*{-10pt}
\tiny
\begin{center}
\def\temptablewidth{1.0\textwidth}
{\rule{\temptablewidth}{1pt}}
\begin{tabular*}{\temptablewidth}{@{\extracolsep{\fill}}ccccccc}
Rank & May 1 & May 10 & May 20 & June 10 & July 6 \\\hline

91	& & & & &													Cambodia	7\\
92	& & & & &													Netherlands Antilles, Sint Maarten	7\\
93	& & & & &													Luxembourg	6\\
94	& & & & &													Algeria	5\\
95		& & & & &												Laos	5\\
96		& & & & &												Nepal	5\\
97		& & & & &												Netherlands, Aruba	5\\
98	& & & & &													Tunisia	5\\
99	& & & & &													U. K., Guernsey, Crown Dependency	5\\
100	& & & & &													France, French Polynesia, FOC 	4\\
101	& & & & &													Iceland	4\\
102	& & & & &													Oman	4\\
103	& & & & &													Cap Verde	3\\
104		& & & & &												Ethiopia	3\\
105	& & & & &													France, Martinique, FOC 	3\\
106	& & & & &													Lithuania	3\\
107	& & & & &													Russia 	3\\
108	& & & & &													Antigua and Barbuda	2\\
109	& & & & &													British Virgin Islands, UKOT 	2\\
110	& & & & &													Cote d'Ivoire	2\\
111	& & & & &													Fiji	2\\
112	& & & & &													France, Guadaloupe, FOC	2\\
113		& & & & &												Guyana	2\\
114		& & & & &												The former Yugoslav Rep. of Macedonia	2\\
115	& & & & &													Vanuatu	2\\
116	& & & & &													Bermuda, UKOT 	1\\
117	& & & & &													Bosnia and Hezegovina	1\\
118		& & & & &												Cook Island	1\\
119		& & & & &												Croatia	1\\
120	& & & & &													Dominica	1\\
121	& & & & &													France Saint Martin, FOC	1\\
122	& & & & &													Iran	1\\
123	& & & & &													Latvia	1\\
124	& & & & &													Libya	1\\
125	& & & & &													Mauritius	1\\
126	& & & & &													Myanmar (Burma)	1\\
127	& & & & &													Palau	1\\
128	& & & & &													Papua New Guinea	1\\
129	& & & & &													Saint Lucia	1\\
130	& & & & &													Samoa	1\\
131	& & & & &													Syria	1\\
132	& & & & &													Uganda	1\\
133	& & & & &													Ukraine	1\\
134	& & & & &													U. K., Isle of Man, Crown Dependency	1\\
135	& & & & &													USA Virgin Islands	1\\
Total: & 367 & 4379 & 10243 & 27737 & 94512\\

       \end{tabular*}
       {\rule{\temptablewidth}{1pt}}
       \end{center}
\end{table}




\end{document}